# GALA-*n*: Generic Architecture of Layout-Aware *n*-Bit Quantum Operators for Cost-Effective Realization on IBM Quantum Computers

Ali Al-Bayaty
*Electrical and Computer Engineering Dept.*
*Portland State University*
Oregon, USA
albayaty@pdx.edu

Marek Perkowski
*Electrical and Computer Engineering Dept.*
*Portland State University*
Oregon, USA
h8mp@pdx.edu

*Abstract*—A generic architecture of *n*-bit quantum operators is proposed for cost-effective transpilation, based on the layouts and the number of *n* neighbor physical qubits for IBM quantum computers, where $n \geq 3$. This proposed architecture is termed "GALA-*n* quantum operator". The GALA-*n* quantum operator is designed using the visual approach of the Bloch sphere, from the visual representations of the rotational quantum operations for IBM native gates ($\sqrt{X}$, X, RZ, and CNOT). In this paper, we also proposed a new formula for the quantum cost, which calculates the total numbers of native gates, SWAP gates, and the depth of the final transpiled quantum circuits. This formula is termed the "transpilation quantum cost". After transpilation, our proposed GALA-*n* quantum operator always has a lower transpilation quantum cost than that of conventional *n*-bit quantum operators, which are mainly constructed from costly *n*-bit Toffoli gates.

*Keywords—Boolean gates, Bloch sphere, Fredkin gate, IBM Brisbane QPU, IBM layout, IBM quantum computer, Miller gate, native gates, Phase gates, quantum computing, Toffoli gate, transpilation quantum cost*

## 1 INTRODUCTION

A quantum gate, in general, is classified into a single-qubit, double-qubit, or multiple-qubit gate. A single-qubit gate directly performs its quantum operation on its only qubit. A double-qubit gate performs its quantum operation on two qubits, and a multiple-qubit gate performs its quantum operation on *n* qubits, where $n \geq 3$. For a double-qubit gate, one of the qubits controls the quantum operation of such a gate onto the other qubit as the outcome; hence, the former qubit is termed the "control qubit" and the latter qubit is termed the "target qubit". Consequently, a multiple-qubit gate has $n - 1$ control qubits and one target qubit. To perform the quantum operation of a double-qubit or multiple-qubit gate on the target qubit, all control qubits should be set to the state of $|1\rangle$, except for the following two quantum gates.

1. The SWAP gate (as a double-qubit gate) has two target qubits and has no control qubits specified, and the SWAP gate always switches the indices of two input qubits as swapped target qubits.
2. The controlled-SWAP or Fredkin gate (as a multiple-qubit gate) has one control qubit and two target qubits, and the Fredkin gate performs the same switching operation as the SWAP gate when the control qubit is only set to the state of $|1\rangle$.

The operation of a quantum gate (or a set of quantum gates) can be categorized into a Boolean or Phase operator, which is an operational component of a Boolean or Phase oracle [1-5], respectively. The quantum Boolean operator acts as the quantum counterpart of the classical Boolean gate (or a set of classical Boolean gates), such as AND, NAND, OR, NOR, XOR, XNOR, implication, and inhibition [6, 7]. Hence, the outcome of a quantum Boolean operator results in flipping the state of a qubit from $|0\rangle$ to $|1\rangle$ or vice versa, depending on the purpose of a quantum operation. In contrast, the quantum Phase operation operates on the phase of a qubit using a quantum gate (or a set of quantum gates), such as controlled-Z (CZ) and multi-controlled-Z (MCZ) gates. Therefore, the outcome of a quantum Phase operator results in rotating the phase of a qubit from $+|x\rangle$ to $-|x\rangle$ or vice versa, depending on the purpose of a quantum operation, where $|x\rangle$ is the state of a qubit.

The Boolean or Phase quantum operator can be represented either (i) mathematically using the unitary matrices [2, 4, 5] that numerically represent their quantum gates, or (ii) geometrically using the Bloch sphere [2, 4, 5] that visually demonstrates the states' transitions of qubits after applying quantum gates.

To realize a practical quantum application, a quantum circuit (as a set of Boolean and Phase quantum operators) is synthesized into a quantum computer, i.e., a quantum processing unit (QPU). In IBM terminologies, the process of synthesis is termed "transpilation" [2, 8]. The transpilation process is the decomposition of a quantum circuit into a set of native gates, which are the so-called "basis gates". These native gates are routed, translated, and then appropriately mapped into the architecture (the layout) of an IBM QPU. In this paper, the `ibm_brisbane` QPU of 127 qubits was utilized to transpile different quantum circuits. Based on the technical specifications of this IBM QPU [9], the single-qubit gates (I, $\sqrt{X}$, X, and RZ($\theta$)) and the double-qubit gate (CNOT or ECR [10, 11]) are IBM native gates, where $\theta$ is an arbitrary rotational angle in radians. If a quantum circuit is constructed from a set of non-native single-qubit and double-qubit gates, then such gates need to be transpiled to the set of aforementioned native gates. However, all multiple-qubit gates, such as $n$-bit Toffoli gate, always need to be transpiled to a sequence of native gates for all IBM QPUs, where $n \geq 3$ qubits.

In our work [2], we proposed the layout-aware $n$-bit Toffoli gate as a cost-effective multiple-qubit gate realization for different IBM QPUs. This layout-aware $n$-bit Toffoli gate was geometrically designed and mathematically modeled using the Bloch sphere approach. After transpilation using an IBM QPU, our proposed gate always has a lower transpilation quantum cost (TQC [2]) than that of the conventional $n$-bit Toffoli gate, due to the fewer utilized numbers of native gates and SWAP gates with a lower depth, where $3 \leq n \leq 7$ qubits. The depth of a quantum circuit is the critical longest path through the total number of native gates and SWAP gates; such that, a higher depth produces a longer delay for the final transpiled quantum circuit, which increases the decoherence time for the physical qubits of an IBM QPU [12-14].

In this paper, the layout-aware $n$-bit Toffoli gate is considered as the generic block of building many quantum Boolean operators (AND, NAND, OR, NOR, implication, and inhibition) and quantum Phase operators (MCZ) using the Bloch sphere approach, and a generic architecture of layout-aware $n$-bit quantum Boolean and Phase operators is proposed for IBM QPUs. Different experiments for this generic architecture of different operators were transpiled and evaluated using `ibm_brisbane` QPU of 127 qubits. After transpilation using this QPU, it was observed that our proposed generic architecture always has a lower TQC than that of the conventional designs of Boolean and Phase operators. All quantum operators are transpiled using the standard IBM transpilation library (Transpiler) [8].

The generic architecture of layout-aware $n$-bit quantum Boolean and Phase operators is designed to be utilized as a transpilation software package for IBM quantum systems. For instance, when a designer transpiles a quantum circuit consisting of many conventional Boolean and Phase operators, our proposed transpilation software package will transpile these conventional operators to their equivalent layout-aware $n$-bit quantum Boolean and Phase operators, for a lower TQC based on the layout of an IBM QPU.

Since the conventional $n$-bit Toffoli gate is mostly associated with building different quantum operators, various decomposition and simplification were performed on this non-native multiple-qubit gate. Barenco *et al.* [15] investigated different approaches to decompose the conventional 3-bit Toffoli gate into a set of single-qubit gates (RY) and double-qubit gates (controlled-V and controlled-V$^{\dagger}$). Also, the authors decomposed the conventional $n$-bit Toffoli gate into a set of conventional 3-bit Toffoli gates. Schmitt and De Micheli [16] presented the "tweedledum" as the open-source library that aims to link the high-level algorithms and the physical devices (QPUs), also the authors demonstrated the decomposition of conventional 3-bit Toffoli gate using the approaches of Decision Diagrams and Phase polynomials through the use of single-qubit gates (T and T$^{\dagger}$) and double-qubit gates (CNOT). Shende and Markov [17] decomposed the conventional 3-bit Toffoli gate into several single-qubit gates (H, T, and T$^{\dagger}$) for less than six CNOT gates, as well as the authors decomposed the CNOT gates into the CZ gates. Ralph *et al.* [18] decomposed the conventional 3-bit and 4-bit Toffoli gates into reduced sets of CNOT, CZ, T, and S gates for binary (qubits) and ternary (qutrits) using multi-level optical quantum systems. In addition, the authors showed that linear optical circuits could operate with much higher probabilities of success. Lukac *et al.* [19] proposed a method to map a quantum circuit to the layout of a QPU, by using the circuit interaction graph (CIG) that is reduced to minimize the number of SWAP gates and the path length between the neighbor physical qubits, using different layouts of IBM QPUs (Tenerife, Melbourne, Ruschlikon, and Tokyo). Tan and Cong [20] discussed different layout mappings that transform quantum circuits to meet quantum hardware limitations. Additionally, the authors' two synthesizers (optimal and approximate) are introduced to reduce the depth and the number of SWAP gates for different QPUs (IBM QX2, IBM Melbourne, and Rigetti Aspen-4). From the aforementioned research papers, and after the transpilation using different

IBM QPUs, we observed that all conventional *n*-bit Toffoli gates have higher TQC than that of our proposed layout-aware *n*-bit Toffoli gate [2], since (i) the conventional *n*-bit Toffoli gates are decomposed to non-native gates that always need to be transpiled to the native gates, (ii) there are many inserted SWAP gates among the controls and target qubits of the conventional *n*-bit Toffoli gates, and (iii) the final depth of the transpiled conventional *n*-bit Toffoli gates is dramatically increased as *n* increases.

## 2 GENERIC ARCHITECTURE APPROACH

The quantum state of a qubit can be visualized using the Bloch sphere, which is the three-dimensional geometrical sphere of three axes (X, Y, and Z). The Bloch sphere visualizes a quantum operation (as an axial-rotated operation) in Hilbert space ($\mathcal{H}$), when a quantum gate is applied to a qubit [2, 4, 5]. In our research, since the quantum operations of all IBM native gates are mainly rotated along the X and Z axes of the Bloch sphere, the XY-plane (as the top-view of the Bloch sphere) is utilized to design the generic architecture of layout-aware *n*-bit quantum Boolean and Phase operators. In this XY-plane, a native (single-qubit or double-qubit) gate rotates the state of a qubit by a counterclockwise rotational angle ($+ \theta$) or a clockwise rotational angle ($- \theta$), in radians. Figure 1(a) demonstrates the Bloch sphere with its corresponding three axes and their respective axial rotations, and Figure 1(b) illustrates the XY-plane of the Bloch sphere. Note that, in Figure 1, no quantum operation is applied to a qubit yet.

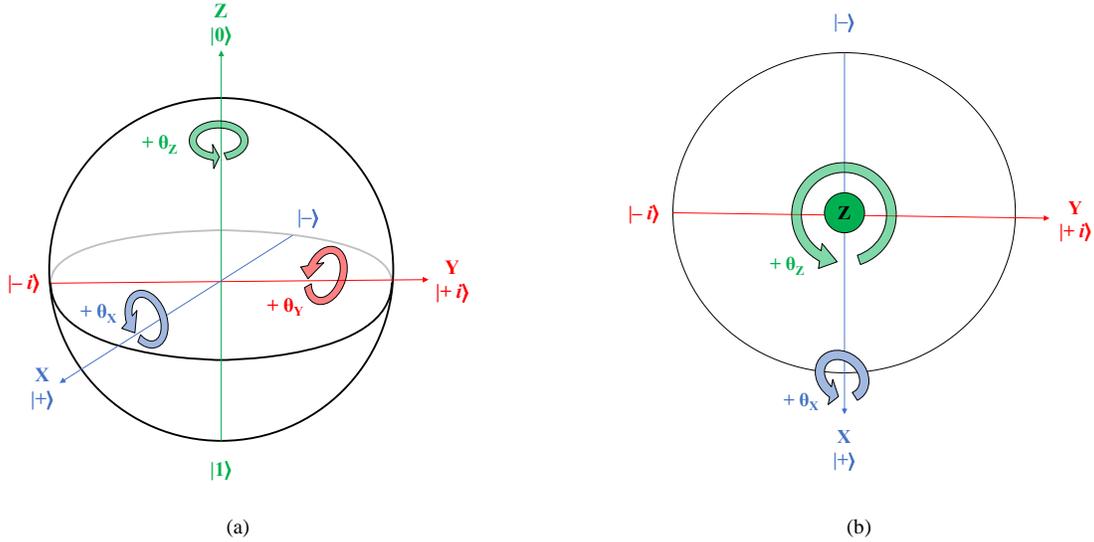

Figure 1. Schematic of the Bloch sphere consisting of (a) the three axes (X in blue, Y in red, and Z in green) with their respective counterclockwise rotational angles ($+\theta_x$, $+\theta_y$, and $+\theta_z$), and (b) the XY-plane (as the top-view) for demonstrating the quantum rotational operations of IBM native (single-qubit and double-qubit) gates around the X and Z axes of the Bloch sphere [2].

In this paper, Dirac notation [4, 5] will be employed for (i) the Bloch sphere visualization, (ii) the state's transitions of qubits, and (iii) the design of the generic architecture of layout-aware *n*-bit quantum Boolean and Phase operators. All IBM native (single-qubit and double-qubit) gates are unitary gates [4, 5], and a few of them are non-Hermitian gates [4, 5], i.e., their quantum operations are not their own inverses. For instance, the $\sqrt{X}^{\dagger}$ or $RX(-\pi/2)$ is the inverse gate for the $\sqrt{X}$ or $RX(+\pi/2)$ gate, the $S^{\dagger}$ or $RZ(-\pi/2)$ is the inverse gate for the S or $RZ(+\pi/2)$ gate, and the $T^{\dagger}$ or $RZ(-\pi/4)$ is the inverse gate for the T or $RZ(+\pi/4)$ gate. Note that some rotational gates alter the state's phase of a qubit; such that, the choice of rotational gates is a critical factor in the design of a quantum circuit. For instance, the Z gate (as the rotation of $+\pi$ in radians) is not the same as the $RZ(+\pi)$ gate, due to the global phase differences; such that, $RZ(+\pi) = -iZ$.

## 2.1 Layout-Aware *n*-Bit Toffoli Gate

In our work [2], the layout-aware *n*-bit Toffoli gate was geometrically designed, mathematically modeled, and algorithmically implemented as a quantum circuit of straight mirrored hierarchical structure, as stated therein in Definition 1 and Definition 2. Such that:

- The straight structure eliminates the SWAP gates among the controls and target qubits, by applying all native gates to the target qubit based on the layout of an IBM QPU; thus, the complexity of mapping the final transpiled quantum circuit into an IBM QPU is decreased.
- The mirrored structure ensures the symmetrical construction of a quantum circuit; thus, when any of the controls is in the state of $|0\rangle$, then the quantum circuit is self-destructive and the target qubit remains at its initial state of $|0\rangle$.
- The hierarchical structure maintains the generation of layout-aware *n*-bit Toffoli gate from the repetitions of the layout-aware *x*-bit Toffoli gates, where $3 \leq x \leq n-1$; thus, the layout-aware *n*-bit Toffoli gate is hierarchically generated from the layout-aware $(n-1)$-bit Toffoli gate, then the layout-aware $(n-2)$-bit Toffoli gate, and so on until the layout-aware 3-bit Toffoli gate.

For the layout-aware *n*-bit Toffoli gate, the target qubit is initially set to the state of $|0\rangle$. To utilize the XY-plane of the Bloch sphere, two Hadamard (H) gates are employed. The first H gate places the target qubit onto the XY-plane, as the first quantum superposition operation. While the second H gate places the target qubit on either $|0\rangle$ or $|1\rangle$ state, as the second quantum superposition operation. The second H gate is applied after a set of axial rotations around the XY-plane depending on all states of control qubits. Such that, the second H gate is intended to find the final solution as a false value ($|0\rangle$ state) or a true value ($|1\rangle$ state), along the Z-axis of the Bloch sphere for the final classical measurement, as shown in Figure 2. The H gate is IBM non-native gate, and it should be decomposed to the sequence $\{RZ(+\pi/2) \sqrt{X} RZ(+\pi/2)\}$ of IBM native gates. For the purpose of simple illustrations, the H gates are kept in the forthcoming figures in this paper; however, they should be replaced by IBM native gates before transpilation.

Figure 2(a) demonstrates the straight mirrored hierarchical quantum circuit of the layout-aware 3-bit Toffoli gate, and Figure 2(b) illustrates the geometrical design approach of the layout-aware 3-bit Toffoli gate using the XY-plane of the Bloch sphere, where the target qubit is in the state of $|+\rangle$ after applying the first H gate and both control qubits ($C_0$ and $C_1$) are set to the state of $|1\rangle$. Finally, Figure 3 depicts the straight mirrored hierarchical structure of the layout-aware *n*-bit Toffoli gate. In our work [2], we proved that the layout-aware *n*-bit Toffoli gate is the quantum counterpart of the classical Boolean *n*-bit AND gate. For a lower transpilation quantum cost (TQC), the target qubit can be switched with any of the control qubits, for cost-effective mapping based on the layout of an IBM QPU.

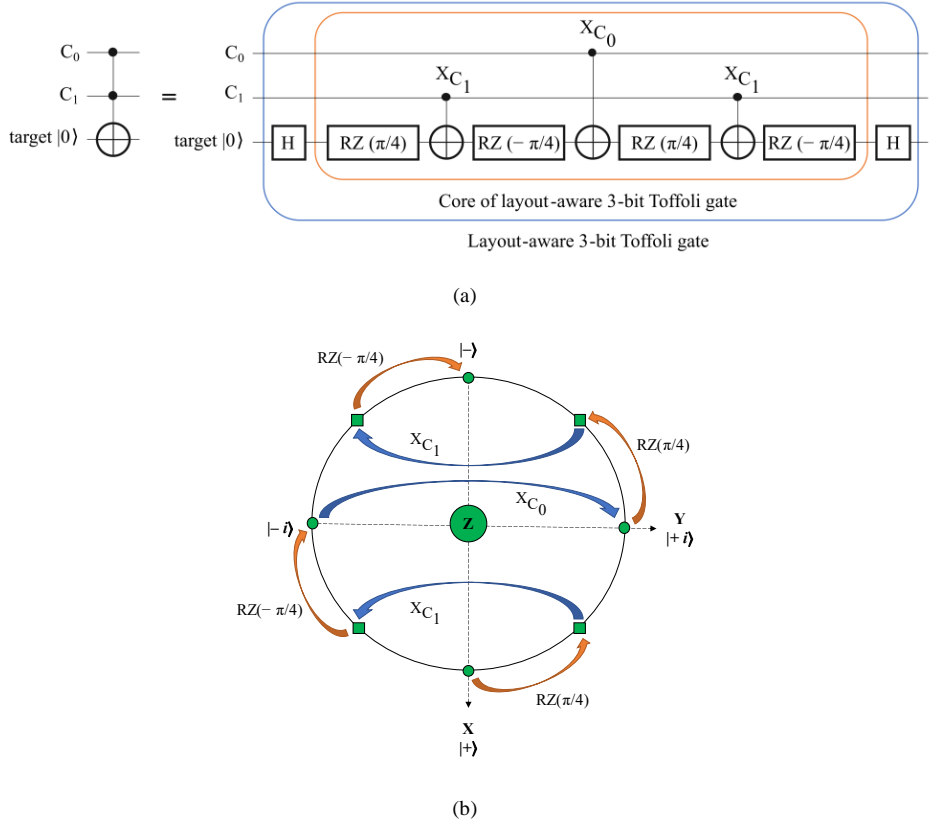

Figure 2. The layout-aware 3-bit Toffoli gate: (a) the construction of its straight mirrored hierarchical quantum circuit, where the rotational angles $\theta = \pm \pi/4$ radians for the four RZ gates, and (b) the geometrical representation of the XY-plane of the Bloch sphere for rotating the target qubit (of the $|+\rangle$ state after applying the first H gate). When both controls ($C_0$ and $C_1$) are set to the states of $|1\rangle$, the target qubit will be in the $|-\rangle$ state, and the second H gate transforms it to the state of $|1\rangle$ to represent the true value of a solution [2].

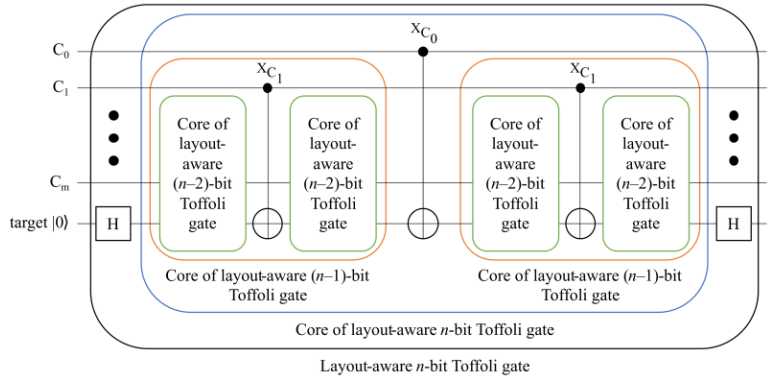

Figure 3. Schematic of the straight mirrored hierarchical quantum circuit consisting of: (i) four cores of layout-aware ($n$–2)-bit Toffoli gate (in green areas), (ii) two cores of layout-aware ($n$–1)-bit Toffoli gate (in orange areas), (iii) one core of layout-aware $n$-bit Toffoli gate (in the blue area), and (iv) the layout-aware $n$-bit Toffoli gate (in the black area). The layout-aware $n$-bit Toffoli gate has $2^{(n-3)}$ cores of layout-aware 3-bit Toffoli gate plus two H gates, where the rotational angles $\theta = \pm \pi / 2^m$ radians for all RZ gates, $m$ is the total number of control qubits, $m = n - 1$, and $n \geq 3$ qubits [2].

## 2.2 Layout-Aware *n*-Bit Quantum Operators

As a generic architecture approach based on the layout of an IBM QPU, the layout-aware *n*-bit quantum Boolean and Phase operators are influenced by the design of layout-aware *n*-bit Toffoli gate [2] and the symmetrical circuits construction [18, 21, 22], where $n \geq 3$ qubits. Hence, the layout-aware *n*-bit quantum Boolean and Phase operators are designed by inheriting the same methodologies of using (i) the straight mirrored hierarchical quantum circuit construction, (ii) the XY-plane of the Bloch sphere approach, (iii) the IBM native (single-qubit and double-qubit) gates, and (iv) the switchable target qubit with any of the control qubits for cost-effective transpilation. The introduced quantum gates, in [18, 21, 22], may not fully utilize the IBM native gates. However, in our proposed generic architecture approach, all utilized quantum gates are IBM native gates. Such that, this generic architecture of layout-aware *n*-bit quantum operator always has lower TQC after transpilation using an IBM QPU.

For ease of description and use, our proposed generic architecture of layout-aware *n*-bit quantum operator is abbreviated as "GALA-*n* quantum operator", for both Boolean and Phase operations, where *n* is the total number of controls and target qubits. The GALA-*n* quantum operator consists of an internal quantum circuit, two superposition gates ($SP_1$ and $SP_2$), and two auxiliary gates ($AX_1$ and $AX_2$). The internal quantum circuit is designed to perform either a Boolean or a Phase operation based on defined rotational angles ($\theta$), and this internal quantum circuit is termed "$Core_n (\theta)$".

***Definition 1.*** *GALA-n quantum operator is the generic architecture of layout-aware n-bit quantum operator, for both Boolean and Phase operations, where $n \geq 3$ qubits. GALA-n quantum operator is a quantum circuit of straight mirrored hierarchical structure for $n - 1$ input qubits (controls) and one out qubit (target). This quantum circuit is only constructed from IBM native single-qubit gates (I, $\sqrt{X}$, X, and RZ($\pm \theta$)) and IBM native double-qubit gates (CNOT or ECR), to ensure the cost-effective transpilation into the layout of an IBM QPU, where $\pm \theta$ is the rotational angle in radians ▫*

***Definition 2.*** *$Core_n (\theta)$ is the internal quantum circuit of the GALA-n quantum operator. This internal quantum circuit is designed to perform either Boolean or Phase operation based on defined rotational angles ($\theta$), where $n \geq 3$ qubits and $\theta$ is a set of rotational angles in radians. $Core_n (\theta)$ is only constructed from IBM native single-qubit gates (RZ($\pm \theta$)) and IBM native double-qubit gates (CNOT or ECR), to ensure the cost-effective transpilation into the layout of an IBM QPU ▫*

To construct a GALA-*n* quantum operator, the GALA-3 quantum operator needs to be constructed first. The GALA-3 quantum operator consists of two control qubits (termed as "$in_0$" and "$in_1$"), one target qubit (termed as "out"), $Core_3 (\theta)$, two superposition gates ($SP_1$ and $SP_2$), and two auxiliary gates ($AX_1$ and $AX_2$), as stated in (1) and (2) below. Figure 4 illustrates the straight mirrored hierarchical quantum circuit of the GALA-3 quantum operator.

$$\text{GALA-3 quantum operator} = SP_1 \ AX_1 \ \{Core_3 (\theta)\} \ AX_2 \ SP_2 \qquad (1)$$

$$Core_3 (\theta) = RZ_1 \ X_{in_1} \ RZ_2 \ X_{in_0} \ RZ_3 \ X_{in_1} \ RZ_4 \qquad (2)$$

Where,
$SP_1$ and $SP_2$ are superposition native single-qubit gates, such as the decomposed H, $\sqrt{X}$, and the decomposed $\sqrt{X}^\dagger$,
$AX_1$ and $AX_2$ are auxiliary native single-qubit gates for different quantum Boolean and Phase operations,
$RZ_1$, $RZ_2$, $RZ_3$, and $RZ_4$ are native single-qubit RZ gates of angles ($\theta_1$, $\theta_2$, $\theta_3$, and $\theta_4$) in radians, respectively, and
$X_{in_0}$ and $X_{in_1}$ are native double-qubit CNOT gates that exist when input qubits ($in_0$ and $in_1$) are set to the $|1\rangle$ state, respectively.

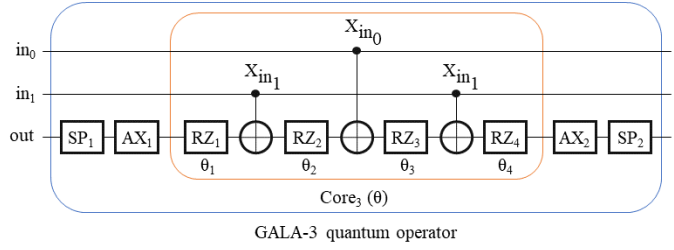

GALA-3 quantum operator

Figure 4. The generic architecture of layout-aware 3-bit quantum (Boolean or Phase) operator (as GALA-3 in blue area) consisting of: (i) straight mirrored hierarchical structure, (ii) core of GALA-3 quantum operator (as $Core_3(\theta)$ in orange area), (iii) two superposition gates ($SP_1$ and $SP_2$), and (iv) two auxiliary gates ($AX_1$ and $AX_2$). $Core_3(\theta)$ has four RZ gates of rotational angles $\theta$ ($\theta_1, \theta_2, \theta_3$, and $\theta_4$), and three CNOT gates. This GALA-3 quantum operator has two inputs ($in_0$ and $in_1$ as the control qubits) and one output (out as the target qubit). The out qubit is switchable with any of the input qubits for cost-effective transpilation. All single-qubit and double-qubit gates are IBM native gates.

To construct the straight mirrored hierarchical structure of GALA-3 quantum operator, the main concept of designing $Core_3(\theta)$ is geometrically based upon the applied rotational gates (RZ and CNOT) on the out qubit, as the visual rotational reflections on the XY-plane of the Block sphere when $\theta$ ($\theta_1, \theta_2, \theta_3$, and $\theta_4$) are accurately chosen for $\pm \pi/4$ in radians, as stated in the following steps and shown in Figure 5.

1. The left side of $Core_3(\theta)$ rotates the out qubit in the lower half of XY-plane, when the input qubit ($in_1$) is set to the $|1\rangle$ state and the rotational angles ($\theta_1$ and $\theta_2$) of $RZ_1$ and $RZ_2$ gates cover the entire area of this lower half.
2. The center of $Core_3(\theta)$ rotates the out qubit around the equator of XY-plane, when the input qubit ($in_0$) is set to the $|1\rangle$ state.
3. The right side of $Core_3(\theta)$ rotates the out qubit in the upper half of XY-plane, when the input qubit ($in_1$) is set to the $|1\rangle$ state and the rotational angles ($\theta_3$ and $\theta_4$) of $RZ_3$ and $RZ_4$ gates cover the entire area of this upper half.

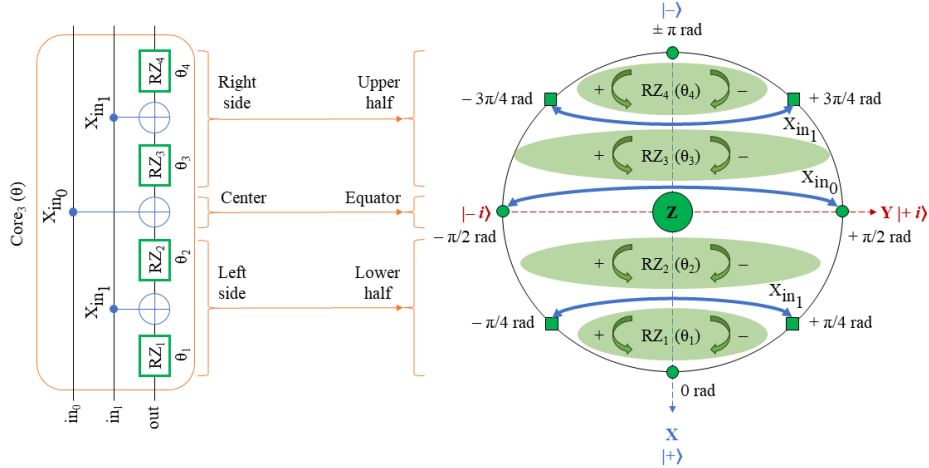

Figure 5. The geometrical concept of designing $Core_3(\theta)$ of the GALA-3 quantum operator (on the left), by using the visual rotational reflections on the XY-plane of the Block sphere (on the right). After applying the rotational gates (RZ and CNOT) on the out qubit: (i) the CNOT gates and their respective arrows (in blue color) indicate the quantum rotations around the X-axis of the Bloch sphere in $\pm \pi$ rad, (ii) the RZ gates and their respective arrows (in green color) indicate the quantum rotations around the Z-axis of the Bloch sphere for $\theta$ ($\theta_1, \theta_2, \theta_3$, and $\theta_4$) in $\pm \pi/4$ rad, (iii) the left side of $Core_3(\theta)$ performs the quantum rotations to cover the entire area of the lower half of XY-plane, (iv) the center of $Core_3(\theta)$ performs the axial quantum rotation around the equator of XY-plane, (v) the right side of $Core_3(\theta)$ performs the quantum rotations to cover the entire area of the upper half of XY-plane, and (vi) all quantum rotations (as quantum gates) are only applied to the out qubit of GALA-3 quantum operator when both inputs ($in_0$ and $in_1$) are set to the state of $|1\rangle$. All single-qubit and double-qubit gates are IBM native gates.

As shown in Figure 5, the angular rotations (around the Z-axis in radians) start from the X-axis ($|+\rangle$) of XY-plane of the Bloch sphere, by assuming that the out qubit of Core$_3$ ($\theta$) is initially set to $|0\rangle$ state, and SP$_1$ (as H gate) is then applied to this out qubit; otherwise,

1. if the out qubit is initially set to the $|1\rangle$ state and SP$_1$ (as H gate) is applied, then the angular rotations start from the X-axis ($|-\rangle$) of the XY-plane,
2. if the out qubit is initially set to the $|0\rangle$ state and SP$_1$ (as $\sqrt{X}$ gate) is applied, then the angular rotations start from the Y-axis ($|-i\rangle$) of the XY-plane,
3. if the out qubit is initially set to the $|1\rangle$ state and SP$_1$ (as $\sqrt{X}$ gate) is applied, then the angular rotations start from the Y-axis ($|+i\rangle$) of the XY-plane,
4. if the out qubit is initially set to the $|0\rangle$ state and SP$_1$ (as $\sqrt{X}^\dagger$ gate) is applied, then the angular rotations start from the Y-axis ($|+i\rangle$) of the XY-plane, or
5. if the out qubit is initially set to the $|1\rangle$ state and SP$_1$ (as $\sqrt{X}^\dagger$ gate) is applied, then the angular rotations start from the Y-axis ($|-i\rangle$) of the XY-plane.

In our research, the GALA-$n$ quantum operators are divided into GALA-$n$ Boolean operators (AND, NAND, OR, NOR, implication, and inhibition) and GALA-$n$ Phase operator (MCZ). On the one hand, GALA-$n$ Boolean operators can be used to effectively build Boolean oracles, especially for the functions and problems that are structured in Product-of-Sums (POS) [23], Sum-of-Products (SOP) [24], Exclusive-or Sum-of-Products (ESOP) [25-27], constraints satisfiable problems – satisfiability (CSP–SAT) [28, 29], as well as in designing XOR functions of two product gates of different Hamming distances [27, 30], just to name a few. On the other hand, the GALA-$n$ Phase operator can be considered a cost-effective replacement for the conventional non-native MCZ gate, which is widely utilized in building Phase oracles and in constructing Grover diffusion operator [1-3]. This paper is considered as an introduction to build cost-effective GALA-$n$ Boolean and Phase operators for different oracles, and as introductory research for future quantum computing enhancements for different layouts of IBM QPUs.

As stated in (1) and (2) above and shown in Figure 5, the accurate rotational angles $\theta$ ($\theta_1$, $\theta_2$, $\theta_3$, and $\theta_4$) of the four RZ gates are $\pm \pi/4$ radians, which cover the entire areas of the lower and the upper halves of XY-plane of the Bloch sphere. Based upon these chosen rotational angles ($\theta$) and the type of selected AX$_2$ as a native gate, Table 1 formulates the construction of GALA-3 Boolean and Phase operators, where the H non-native gates are decomposed to the sequence {RZ($+\pi/2$) $\sqrt{X}$ RZ($+\pi/2$)} of IBM native gates, and I is IBM identity native gate. The X$_{\text{in}_0}$ and X$_{\text{in}_1}$ gates only exist when in$_0$ and in$_1$ are set to the state of $|1\rangle$. For future quantum computing research, the AX$_1$ gate is reserved for developing new types of interesting quantum gates.

Table 1. Specifications of IBM native (single-qubit and double-qubit) gates for constructing GALA-3 Boolean and Phase operators for IBM QPUs, where H → {RZ($+\pi/2$) $\sqrt{X}$ RZ($+\pi/2$)}.

| 3-bit quantum operator | Quantum operation | Initial state of out qubit | SP$_1$ | AX$_1$ | Core$_3$ ($\theta$) | | | | AX$_2$ | SP$_2$ |
|---|---|---|---|---|---|---|---|---|---|---|
| | | | | | RZ$_1$ ($\theta_1$) | RZ$_2$ ($\theta_2$) | RZ$_3$ ($\theta_3$) | RZ$_4$ ($\theta_4$) | | |
| AND (in$_0$ ∧ in$_1$) | Boolean | $|0\rangle$ | H | I | $-\pi/4$ | $+\pi/4$ | $-\pi/4$ | $+\pi/4$ | I | H |
| NAND ¬ (in$_0$ ∧ in$_1$) | | | | | $-\pi/4$ | $+\pi/4$ | $-\pi/4$ | $+\pi/4$ | RZ($-\pi$) | |
| OR (in$_0$ ∨ in$_1$) | | | | | $+\pi/4$ | $+\pi/4$ | $+\pi/4$ | $+\pi/4$ | RZ($+\pi$) | |
| NOR ¬ (in$_0$ ∨ in$_1$) | | | | | $+\pi/4$ | $+\pi/4$ | $+\pi/4$ | $+\pi/4$ | I | |
| Implication (¬ in$_0$ ∨ in$_1$) | | | | | $-\pi/4$ | $-\pi/4$ | $+\pi/4$ | $+\pi/4$ | RZ($-\pi$) | |
| Inhibition ¬ (¬ in$_0$ ∨ in$_1$) | | | | | $-\pi/4$ | $-\pi/4$ | $+\pi/4$ | $+\pi/4$ | I | |
| MCZ (in$_0$ ∧ in$_1$) "Experimental" | Phase | $|0\rangle$ or $|1\rangle$ | | | $+\pi/2$ | $+\pi/2$ | $-\pi/2$ | $-\pi/2$ | I | |

From Table 1, the H gates (in $SP_1$ and $SP_2$) can be replaced with $\sqrt{X}$ and $\sqrt{X}^{\dagger}$ gates, to construct a few GALA-3 Boolean and Phase operators, as stated in Table 2. The $\sqrt{X}^{\dagger}$ is IBM non-native gate that should be decomposed to three IBM native single-qubit gates for the sequence {RZ(+ π) $\sqrt{X}$ RZ(+ π)}.

Table 2. Specifications of IBM native (single-qubit and double-qubit) gates for constructing GALA-3 Boolean and Phase operators for IBM QPUs, where $\sqrt{X}^{\dagger} \rightarrow$ {RZ(+ π) $\sqrt{X}$ RZ(+ π)}.

| 3-bit quantum operator | Quantum operation | Initial state of out qubit | $SP_1$ | $AX_1$ | $Core_3 (\theta)$ | | | | $AX_2$ | $SP_2$ |
|---|---|---|---|---|---|---|---|---|---|---|
| | | | | | $RZ_1 (\theta_1)$ | $RZ_2 (\theta_2)$ | $RZ_3 (\theta_3)$ | $RZ_4 (\theta_4)$ | | |
| AND ($in_0 \wedge in_1$) | Boolean | $\|0\rangle$ | $\sqrt{X}$ | I | + π/4 | + π/4 | − π/4 | − π/4 | I | $\sqrt{X}^{\dagger}$ |
| NAND ¬ ($in_0 \wedge in_1$) | | | $\sqrt{X}$ | | + π/4 | + π/4 | − π/4 | − π/4 | RZ(− π) | $\sqrt{X}^{\dagger}$ |
| OR ($in_0 \vee in_1$) "FAILED" | | | $\sqrt{X}$ | | + π/4 | + π/4 | + π/4 | + π/4 | RZ(+ π) | $\sqrt{X}^{\dagger}$ |
| NOR ¬ ($in_0 \vee in_1$) "FAILED" | | | $\sqrt{X}$ | | + π/4 | + π/4 | + π/4 | + π/4 | I | $\sqrt{X}^{\dagger}$ |
| Implication (¬ $in_0 \vee in_1$) | | | $\sqrt{X}$ | | + π/4 | − π/4 | + π/4 | − π/4 | RZ(− π) | $\sqrt{X}^{\dagger}$ |
| Inhibition ¬ (¬ $in_0 \vee in_1$) | | | $\sqrt{X}$ | | + π/4 | − π/4 | + π/4 | − π/4 | I | $\sqrt{X}^{\dagger}$ |
| MCZ ($in_0 \wedge in_1$) "FAILED" | Phase | $\|0\rangle$ or $\|1\rangle$ | $\sqrt{X}$ | | + π/2 | + π/2 | − π/2 | − π/2 | I | $\sqrt{X}^{\dagger}$ |

To reflect the correct quantum rotations in the XY-plane of the Bloch sphere, as stated in Table 1 and Table 2, the following key points need to be cautiously undertaken when constructing a single GALA-3 Boolean or Phase operator:

- The rotational angles (θ) of the four RZ gates need to be accurately calculated to cover the entire area of the XY-plane of the Bloch sphere, when the inputs ($in_0$ and $in_1$) are set to the state of $|1\rangle$ for different configurations of $SP_1$ and $SP_2$.
- Neighbor RZ gates are conditionally switchable gates; such that, $RZ_1$ is only switched with $RZ_2$, if and only if $RZ_3$ switches with $RZ_4$, and vice versa, for different configurations of $SP_1$ and $SP_2$.
- The $\sqrt{X}$ and $\sqrt{X}^{\dagger}$ gates can be switched between $SP_1$ and $SP_2$ in different configurations; such that, there will be four different permutative configurations for states superposition on the XY-plane of the Bloch sphere.
- The parametric selection and modification of the rotational angles (θ) of RZ gates, the superposition gates ($SP_1$ and $SP_2$), and the auxiliary gates ($AX_1$ and $AX_2$) are open topics for future quantum computing research, to develop interesting and cost-effective quantum operators and gates based on the layouts and the number of *n* neighbor physical qubits of IBM QPUs.

For ease of illustration and implementation, in this paper, our proposed GALA-*n* based approach is restricted to utilizing the H gates in $SP_1$ and $SP_2$. Based on the chosen θ of RZ gates and the specified type of $AX_2$ gate from Table 1, Figure 6 visualizes the q-spheres [31] for the outcomes of GALA-3 Boolean operators, and Figure 7 visualizes the q-spheres of outcomes for GALA-3 Phase operator. The q-sphere visualizes the global view of multiple qubits on their computational basis states, with their nodes' size as the amplitude probabilities and their colors as the global phases. For all qubits in these q-spheres, as shown in Figure 5 and Figure 6, their Dirac notations are in the form of |out $in_1$ $in_0$⟩.

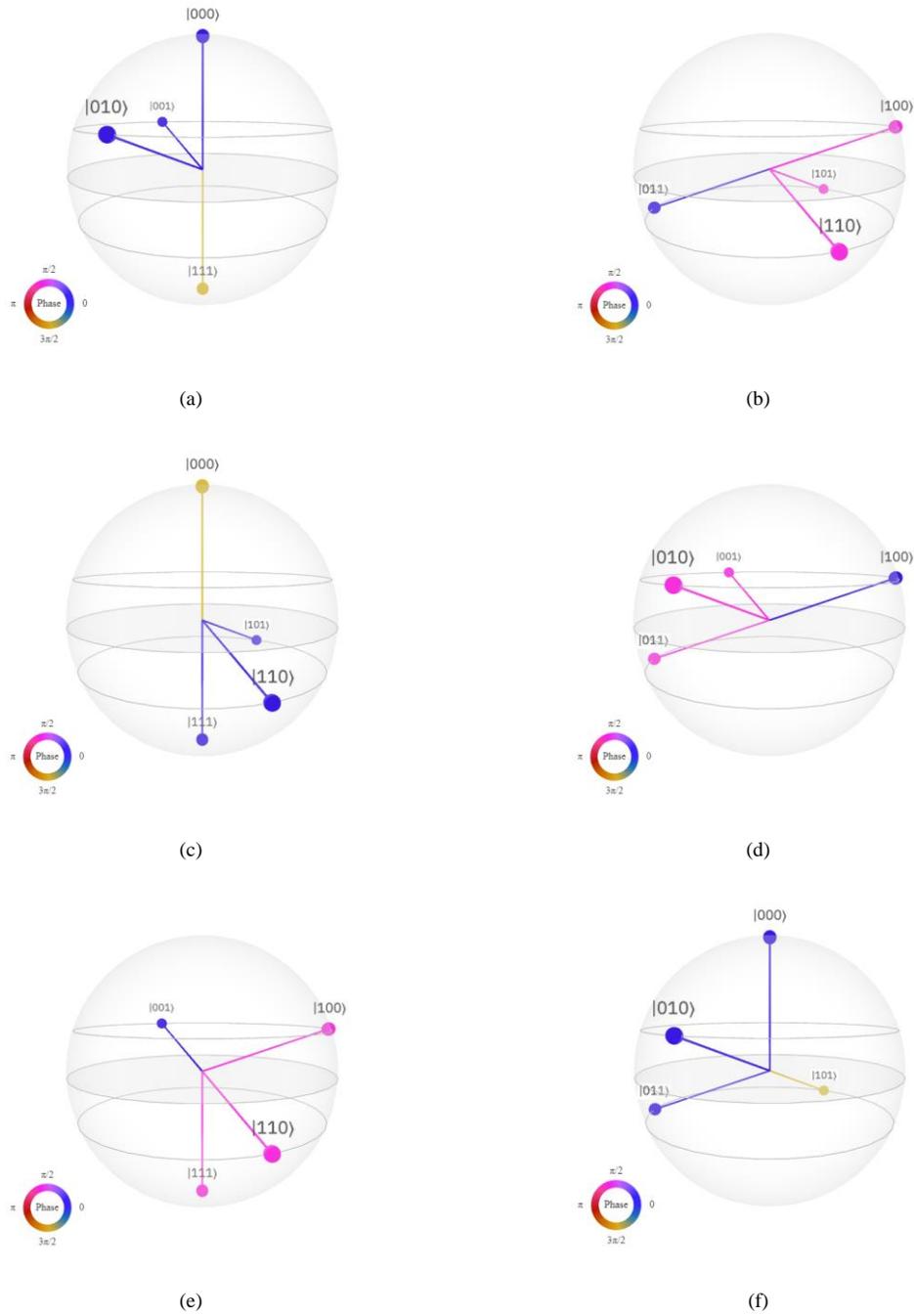

Figure 6. Q-spheres for the outcomes of GALA-3 Boolean operators based on Table 1: (a) AND, (b) NAND, (c) OR, (d) NOR, (e) implication, and (f) inhibition. For all qubits in these q-spheres, their Dirac notations are in the form of |out $in_1$ $in_0$⟩. All out qubits are initially set to state |0⟩. The global phases of 0 radians are indicated by blue color, the global phase shifts of $\pi/2$ radians are indicated by pink color, and the global phase shifts of $3\pi/2$ radians are indicated by yellow color.

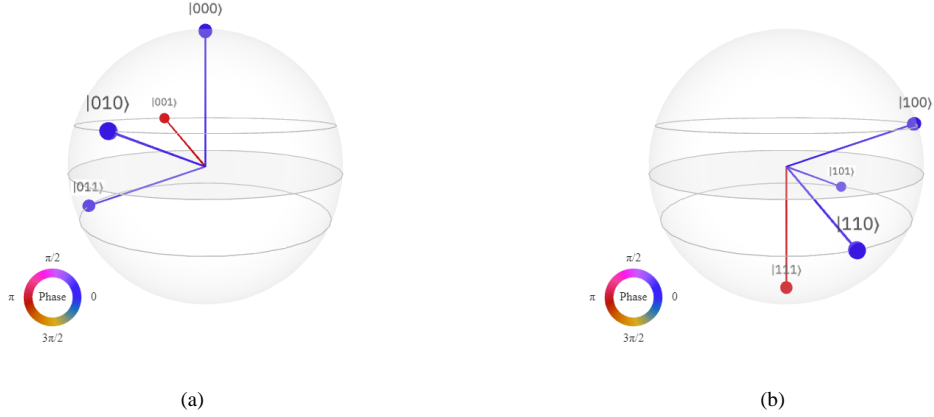

Figure 7. Q-spheres for the outcomes of the GALA-3 Phase operator based on Table 1: (a) MCZ with the out qubit initially set to the $|0\rangle$ state, and (b) MCZ with the out qubit initially set to the $|1\rangle$ state. For all qubits in these q-spheres, their Dirac notations are in the form of $|out\ in_1\ in_0\rangle$. The global phases of 0 radians are indicated by blue color, and the global phase shifts of $\pi$ radians are indicated by red color. The GALA-3 MCZ operator is currently under research and development.

From Table 1 and Figure 6, the global phase shifts for the out qubits are not a critical issue, since these GALA-3 Boolean operators are used to build Boolean oracles that do not mostly utilize the phases inversion [1-3], as well as these global phase shifts will be negligible after the classical measurements. From Table 1 and Figure 7, the GALA-3 MCZ operator is currently under research and development for future Phase-based enhancements.

Based on the approach of using the straight mirrored hierarchical structure, the GALA-4 quantum operator can be constructed by: (i) mirroring $Core_3\ (\theta)$, (ii) inserting the new input qubit as a new CNOT gate between the mirrored cores to form $Core_4\ (\theta)$, and finally (iii) $SP_1$, $AX_1$, $ZX_2$, and $SP_2$ gates are added to form the GALA-4 quantum operator, as stated in (3) and (4) below and illustrated in Figure 8. For ease of illustration, the new input qubit is indexed as "$in_0$" and the input qubits of $Core_3\ (\theta)$ are shifted to "$in_1$" and "$in_2$". The $X_{in_0}$, $X_{in_1}$, $X_{in_2}$ gates only exist when their input qubits are set to the state of $|1\rangle$. The rotational angles ($\theta$) of the eight RZ gates for the mirrored $Core_3\ (\theta)$ are reduced by half ($\pm\ \pi/8$ radians), to cover the entire areas of the lower and the upper halves of the XY-plane of the Bloch sphere. The specifications of $SP_1$, $AX_1$, $AX_2$, and $SP_2$ gates are taken from Table 1, to reflect the same quantum (Boolean or Phase) operation of GALA-3 on GALA-4.

$$\text{GALA-4 quantum operator} = SP_1\ AX_1\ \{Core_4\ (\theta)\}\ AX_2\ SP_2 \qquad (3)$$

$$Core_4\ (\theta) = \{Core_3\ (\theta)\}\ X_{in_0}\ \{Core_3\ (\theta)\} \qquad (4)$$

Figure 9 demonstrates the $Core_4\ (\theta)$ of the GALA-4 quantum operator and the visual rotational reflections on the XY-plane of the Block sphere, based upon the applied rotational gates (RZ and CNOT) on the out qubit for the rotational angles $\theta$ ($\theta_1$, $\theta_2$, $\theta_3$, and $\theta_4$) in $\pm\ \pi/8$ radians as follows.

1. The left side of $Core_4\ (\theta)$, i.e., $Core_3\ (\theta)$, rotates the out qubit in the lower half of XY-plane, when the input qubits ($in_1$ and $in_2$) are set to the $|1\rangle$ state and all rotational angles ($\theta_1$, $\theta_2$, $\theta_3$, and $\theta_4$) of four RZ gates cover the entire area of this lower half.
2. The center of $Core_4\ (\theta)$ rotates the out qubit around the equator of XY-plane, when the input qubit ($in_0$) is set to the $|1\rangle$ state.
3. The right side of $Core_4\ (\theta)$, i.e., $Core_3\ (\theta)$, rotates the out qubit in the upper half of XY-plane, when the input qubits ($in_1$ and $in_2$) are set to the $|1\rangle$ state and all rotational angles ($\theta_1$, $\theta_2$, $\theta_3$, and $\theta_4$) of four RZ gates cover the entire area of this upper half.

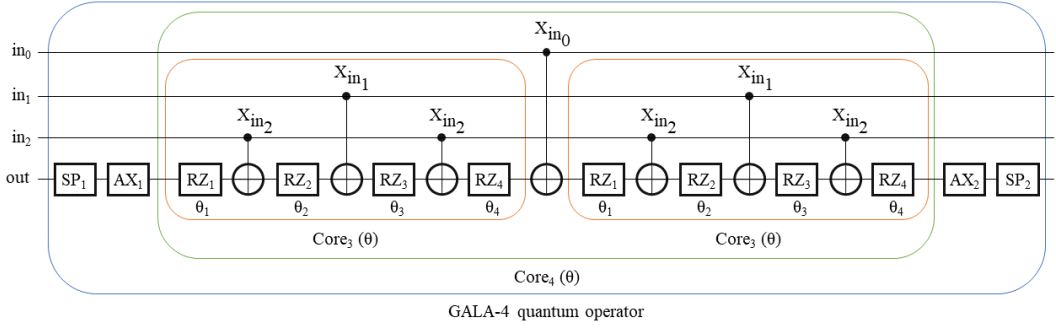

Figure 8. The generic architecture of layout-aware 4-bit quantum (Boolean or Phase) operator (as GALA-4 in blue area) consisting of: (i) straight mirrored hierarchical structure, (ii) one core of GALA-4 quantum operator (as $Core_4$ ($\theta$) in green area), (iii) two cores of GALA-3 quantum operator (as $Core_3$ ($\theta$) in orange areas), (iv) two superposition gates ($SP_1$ and $SP_2$), and (v) two auxiliary gates ($AX_1$ and $AX_2$). The $Core_4$ ($\theta$) has eight RZ gates, seven CNOT gates, and rotational angles $\theta$ ($\theta_1$, $\theta_2$, $\theta_3$, and $\theta_4$) of $\pm \pi/8$ rad. This generic architecture consists of three inputs ($in_0$, $in_1$, and $in_2$ as the control qubits), one output (out as the target qubit), the out qubit is switchable with any of the input qubits for cost-effective transpilation, and all single-qubit and double-qubit gates are IBM native gates.

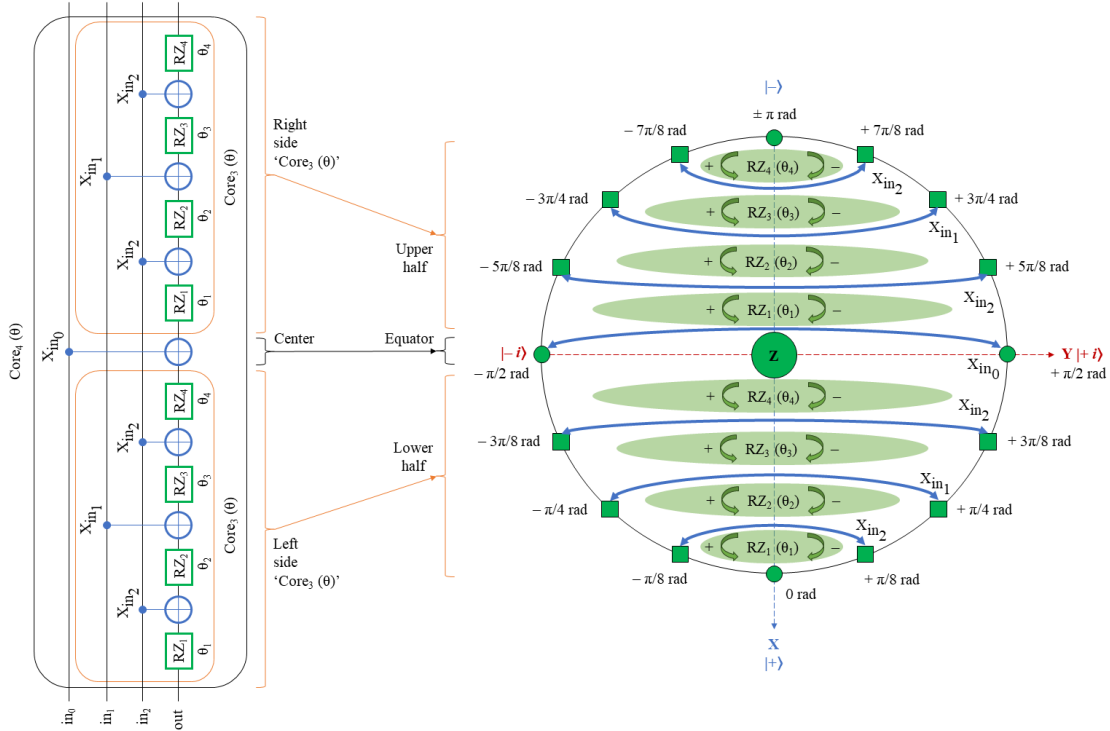

Figure 9. The geometrical concept of designing $Core_4$ ($\theta$) of the GALA-4 quantum operator (on the left), by using the visual rotational reflections on the XY-plane of the Block sphere (on the right). After applying the rotational gates (RZ and CNOT) on the out qubit: (i) the CNOT gates and their respective arrows (in blue color) indicate the quantum rotations around the X-axis of the Bloch sphere in $\pm \pi$ rad, (ii) the RZ gates and their respective arrows (in green color) indicate the quantum rotations around the Z-axis of the Bloch sphere for $\theta$ ($\theta_1$, $\theta_2$, $\theta_3$, and $\theta_4$) in $\pm \pi/8$ rad, (iii) the left side of $Core_4$ ($\theta$), i.e., $Core_3$ ($\theta$), performs the quantum rotations to cover the entire area of the lower half of XY-plane, (iv) the center of $Core_4$ ($\theta$) performs the axial quantum rotation around the equator of XY-plane, and (v) the right side of $Core_4$ ($\theta$), i.e., $Core_3$ ($\theta$), performs the quantum rotations to cover the entire area of the upper half of XY-plane. All quantum rotations (as quantum gates) are only applied to the out qubit of the GALA-4 quantum operator when all inputs ($in_0$, $in_1$, and $in_2$) are set to the state of $|1\rangle$. All single-qubit and double-qubit gates are IBM native gates.

Consequently, for $n \geq 3$ qubits, the GALA-$n$ quantum operator is constructed by mirroring the core of GALA-$(n–1)$ quantum operator, then inserting the new input qubit as a new CNOT gate between the mirrored cores, and SP$_1$, AX$_1$, ZX$_2$, and SP$_2$ gates are added to form the GALA-$n$ quantum operator, as stated in (5) and (6) below and demonstrated in Figure 10. The Core$_n$ ($\theta$) is the internal quantum circuit of the GALA-$n$ quantum operator for the rotational angles ($\theta$) denoted in (7) below, which cover the entire area of the XY-plane of the Bloch sphere when all input qubits are set to the state of $|1\rangle$. The specifications of SP$_1$, AX$_1$, AX$_2$, and SP$_2$ gates are taken from Table 1, to reflect the same quantum (Boolean or Phase) operation of GALA-3 on GALA-$n$.

$$\text{GALA-}n \text{ quantum operator} = \text{SP}_1 \text{ AX}_1 \{\text{Core}_n (\theta)\} \text{ AX}_2 \text{ SP}_2 \tag{5}$$

$$\text{Core}_n (\theta) = \{\text{Core}_{n-1} (\theta)\} \text{ X}_{\text{in}_0} \{\text{Core}_{n-1} (\theta)\} \tag{6}$$

$$\theta = \pm \pi / 2^{n-1} \tag{7}$$

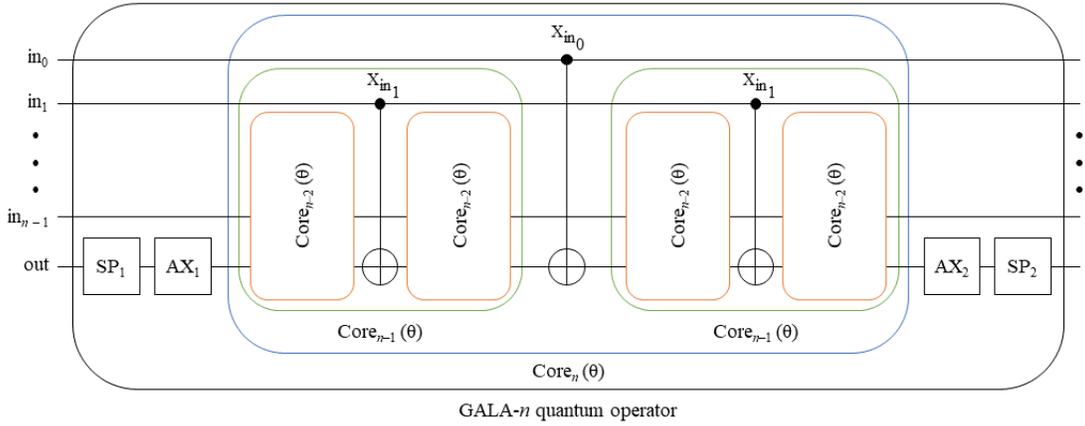

Figure 10. The generic architecture of layout-aware $n$-bit quantum (Boolean or Phase) operator (as GALA-$n$ in black area) consisting of: (i) straight mirrored hierarchical structure, (ii) one core of GALA-$n$ quantum operator (as Core$_n$ ($\theta$) in blue area), (iii) two cores of GALA-$(n–1)$ quantum operator (as Core$_{n-1}$ ($\theta$) in green areas), (iv) four cores of GALA-$(n–2)$ quantum operator (as Core$_{n-2}$ ($\theta$) in orange areas), (v) two superposition gates (SP$_1$ and SP$_2$), and (vi) two auxiliary gates (AX$_1$ and AX$_2$). The GALA-$n$ quantum operator is composed of $2^{(n-3)}$ Core$_3$ ($\theta$) for $\theta$ ($\theta_1$, $\theta_2$, $\theta_3$, and $\theta_4$) = $\pm \pi/2^{n-1}$ rad, and $n \geq 3$ qubits. The out qubit is switchable with any of the input qubits for cost-effective transpilation. All single-qubit and double-qubit gates are IBM native gates.

As stated in (1) till (7) above and shown in Figure 10, the construction of GALA-$n$ quantum (Boolean and Phase) operators can be expressed algorithmically, as stated in ALGORITHM 1, where $n \geq 3$ qubits. The symbol ($\rightarrow$) means "apply to" and the symbol ($\leftarrow$) means "append to". This algorithm constructs the completed straight mirrored hierarchical quantum circuit of the GALA-$n$ quantum operator as the balanced binary tree [32].

| | | |
|---|---|---|
| **ALGORITHM 1: Generic architecture of layout-aware *n*-bit quantum (Boolean or Phase) operator** | | |
| **Inputs:** list of ($n-1$) indices for input qubits, one index for output qubit, and operator type (AND, NAND OR, NOR, implication, inhibition, or MCZ) | | |
| **Output:** quantum circuit (straight mirrored hierarchical structure) of layout-aware *n*-bit quantum operator | | |
| 1. | $\theta = \pm \pi / 2^{(n-1)}$ | ► Rotational angles ($\theta_1, \theta_2, \theta_3$, and $\theta_4$) |
| 2. | **Construct** $Core_3(\theta)$ **of:** | |
| 3. | Two inputs of indices ($n-1$ and $n-2$) | |
| 4. | One output of index ($n$) | |
| 5. | Three CNOT gates | ► From Table 1, from inputs ($n-1; n-2; n-1$) to output ($n$) |
| 6. | Four RZ gates | ► From Table 1, specify $\theta_1, \theta_2, \theta_3$, and $\theta_4$ based on operator type |
| 7. | **End Construct** | |
| 8. | $r = n-3$ | ► Calculate the remaining inputs |
| 9. | **While** ($r > 0$) **do:** | ► Construct $Core_n(\theta)$, as a hierarchical structure |
| 10. | $Core_{n-r+1}(\theta) \leftarrow Core_{n-r}(\theta)$ | ► Attach the previous core, as a straight structure |
| 11. | $Core_{n-r+1}(\theta) \leftarrow CNOT$ | ► Attach Feynman gate, from input ($r$) to output ($n$) |
| 12. | $Core_{n-r+1}(\theta) \leftarrow Core_{n-r}(\theta)$ | ► Attach the previous core, as a mirrored structure |
| 13. | $r = r-1$ | ► Iterate to the next remaining inputs |
| 14. | **End While** | |
| 15. | $SP_1 \rightarrow$ output ($n$) | ► From Table 1, based on operator type and before $Core_n(\theta)$ |
| 16. | $AX_1 \rightarrow$ output ($n$) | ► From Table 1, based on operator type and before $Core_n(\theta)$ |
| 17. | $AX_2 \rightarrow$ output ($n$) | ► From Table 1, based on operator type and after $Core_n(\theta)$ |
| 18. | $SP_2 \rightarrow$ output ($n$) | ► From Table 1, based on operator type and after $Core_n(\theta)$ |

## 2.3 Composing Quantum Gates using GALA-*n* AND Operator

For a lower TQC of the final transpiled quantum circuit into an IBM QPU, we re-designed a set of conventional *n*-bit quantum gates using the GALA-*n* AND operator, where $n \geq 3$ qubits. This set includes *n*-bit controlled-V, *n*-bit controlled-V$^\dagger$, *n*-bit controlled-SWAP (Fredkin), and *n*-bit Miller gates, which are IBM non-native multiple-qubit gates. The *n*-bit controlled-V and *n*-bit controlled-V$^\dagger$ gates are mainly used in constructing *n*-bit Toffoli gates, *n*-bit Peres gates, arbitrary gates for the majority of three variables in a function [15, 21, 33-37], just to name a few. The *n*-bit Fredkin gate is used in generating linear cascade circuits [38, 39]. The *n*-bit Miller gate inverts all input qubits (all in the state of |1⟩) when the out qubit is set to |0⟩ state, and it inverts all input qubits (all in the state of |0⟩) when the out qubit is set to |1⟩ state [33, 37, 40].

The quantum operations of 3-bit controlled-V and 3-bit controlled-V$^\dagger$ gates act similarly as the 3-bit controlled-$\sqrt{X}$ and 3-bit controlled-$\sqrt{X}^\dagger$ non-native gates, respectively. When all input qubits (controls) of 3-bit controlled-V and 3-bit controlled-V$^\dagger$ gates are set to the state of |1⟩, the V ($\sqrt{X}$) and V$^\dagger$ ($\sqrt{X}^\dagger$) gates are then applied to the out qubit (target), as stated in (8) and (9) below when the out qubit is initially set to |0⟩ state, and in (10) and (11) below when the out qubit is initially set to |1⟩ state, respectively.

$$V|0\rangle = \sqrt{X}|0\rangle = \tfrac{1}{\sqrt{2}}\left(|0\rangle + e^{i\,3\pi/2}|1\rangle\right) \qquad (8)$$

$$V^\dagger|0\rangle = \sqrt{X}^\dagger|0\rangle = \tfrac{1}{\sqrt{2}}\left(|0\rangle + e^{i\,\pi/2}|1\rangle\right) \qquad (9)$$

$$V|1\rangle = \sqrt{X}|1\rangle = \tfrac{1}{\sqrt{2}}\left(|0\rangle + e^{i\,\pi/2}|1\rangle\right) \qquad (10)$$

$$V^\dagger|1\rangle = \sqrt{X}^\dagger|1\rangle = \tfrac{1}{\sqrt{2}}\left(|0\rangle + e^{i\,3\pi/2}|1\rangle\right) \qquad (11)$$

Figure 11(a) and Figure 11(b) illustrate our proposed decompositions of the conventional 2-bit controlled-V and the conventional 2-bit controlled-$V^\dagger$ gates, by using IBM native gates and the Bloch sphere approach. Moreover, the H gates are implicitly decomposed to the sequence $\{RZ(+\pi/2)\ \sqrt{X}\ RZ(+\pi/2)\}$ of IBM native gates. From the above two proposed decompositions, the 3-bit controlled-V and 3-bit controlled-$V^\dagger$ gates are then composed using GALA-3 AND operators, instead of using the conventional cost-ineffective 3-bit Toffoli gates, as shown in Figure 12(a) and Figure 12(b), respectively. Note that our aforementioned two proposed decompositions are a cost-effective approach rather than using the standard IBM decompositions, due to the fewer number of utilized IBM native single-qubit and double-qubit gates with lower depth. In other words, our proposed decompositions always have lower TQC for the final transpiled quantum circuit into the layout of an IBM QPU.

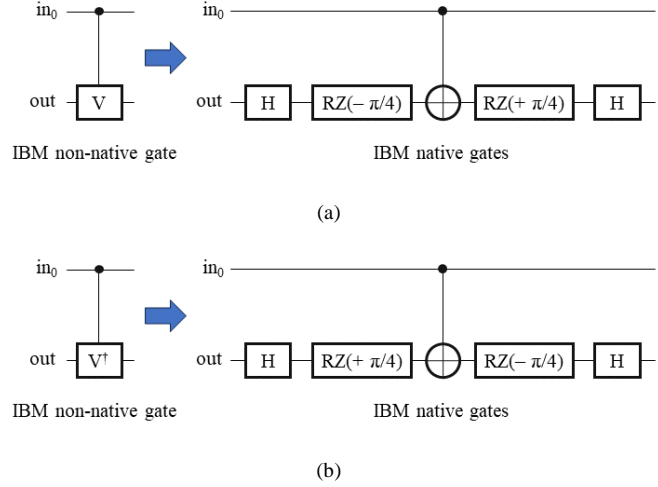

Figure 11. Schematic of proposed cost-effective quantum gates decomposition from IBM non-native gates to IBM native gates: (a) the decomposition of 2-bit controlled-V gate, and (b) the decomposition of 2-bit controlled-$V^\dagger$ gate. The H gates are implicitly decomposed to the sequence $\{RZ(+\pi/2)\ \sqrt{X}\ RZ(+\pi/2)\}$ of IBM native gates.

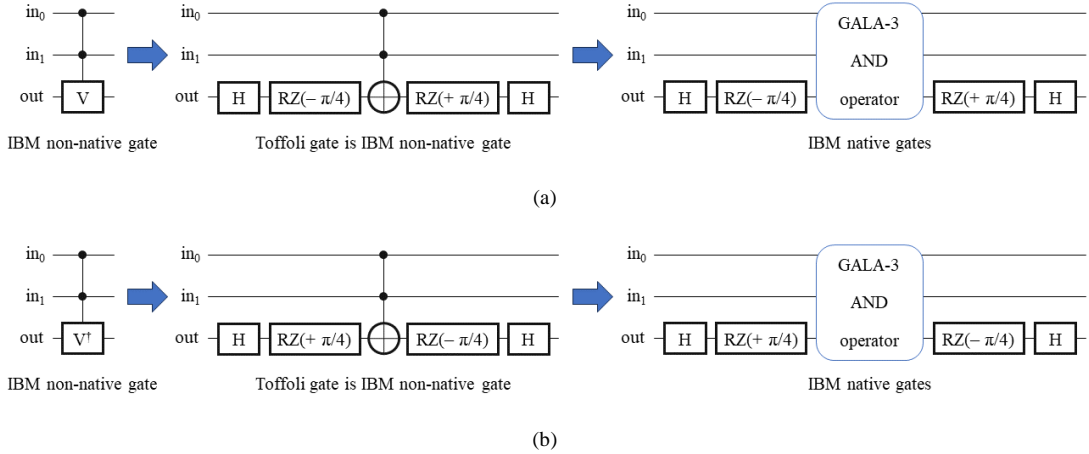

Figure 12. Schematic of proposed cost-effective quantum gates decomposition from IBM non-native gates to IBM native gates using GALA-3 AND operators: (a) the decomposition of 3-bit controlled-V gate, and (b) the decomposition of 3-bit controlled-$V^\dagger$ gate. The H gates are implicitly decomposed to the sequence $\{RZ(+\pi/2)\ \sqrt{X}\ RZ(+\pi/2)\}$ of IBM native gates.

As stated in (8) and (10) above and shown in Figure 12(a), when both input qubits ($in_0$ and $in_1$) are set to the state of $|1\rangle$, Figure 13(a) and Figure 13(b) illustrate the q-spheres for the out qubit initially set to $|0\rangle$ and $|1\rangle$ states, respectively. As stated in (9) and (11) above and as shown in Figure 12(b), when both input qubits ($in_0$ and $in_1$) are set to the state of $|1\rangle$, Figure 13(c) and Figure 13(d) illustrate the q-spheres for the out qubit initially set to $|0\rangle$ and $|1\rangle$ states, respectively. For all qubits in these q-spheres, their Dirac notations are in the form of $|out\ in_1\ in_0\rangle$.

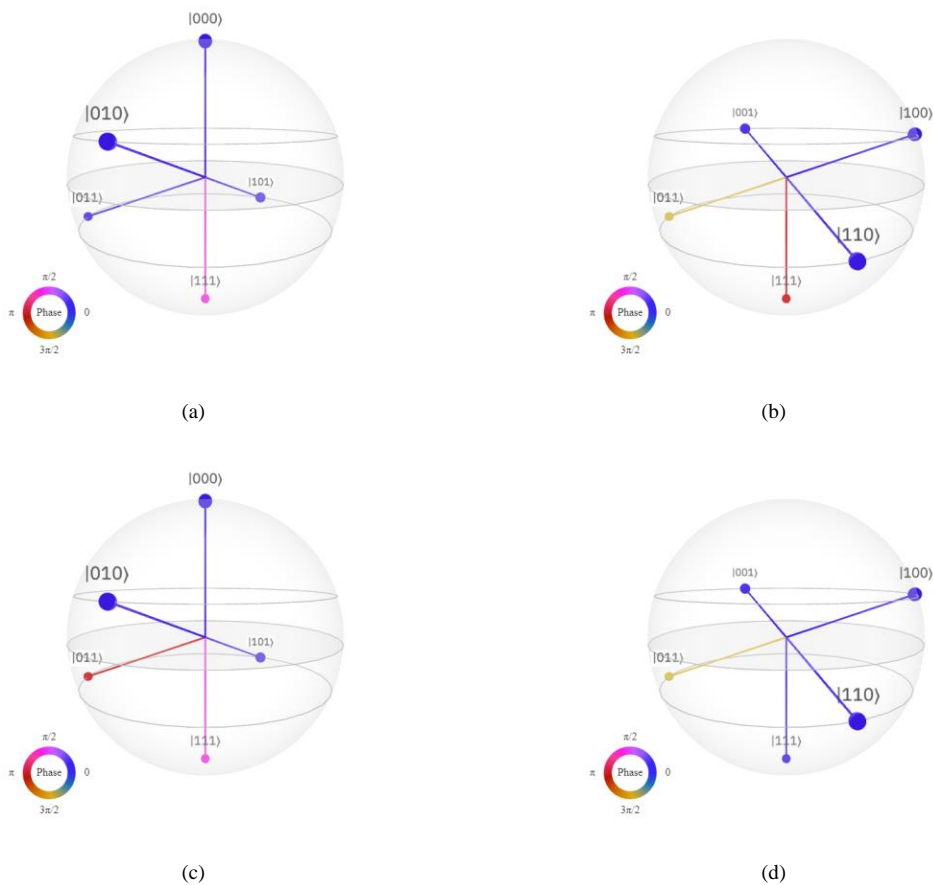

(a)　　　　　　　　　　　　　　　(b)

(c)　　　　　　　　　　　　　　　(d)

Figure 13. Q-spheres for the outcomes of composed 3-bit quantum gates using GALA-3 AND operators: (a) the out initially set to state of $|0\rangle$ for the composed 3-bit controlled-V gate, (b) the out initially set to the state of $|1\rangle$ for the composed 3-bit controlled-V gate, (c) the out initially set to the state of $|0\rangle$ for the composed 3-bit controlled-$V^\dagger$ gate, and (d) the out initially set to the state of $|1\rangle$ for the composed 3-bit controlled-$V^\dagger$ gate. For all qubits in these q-spheres, their Dirac notations are in the form of $|out\ in_1\ in_0\rangle$.

Consequently, the conventional $n$-bit controlled-V and $n$-bit controlled-$V^\dagger$ gates are composed using GALA-$n$ AND operators, as shown in Figure 14(a) and Figure 14(b), respectively. Similarly, from the proposed composition mechanisms of the conventional $n$-bit controlled-V and $n$-bit controlled-$V^\dagger$ gates, the conventional $n$-bit Fredkin and the conventional $n$-bit Miller gates are composed using GALA-$n$ AND operators, as shown in Figure 15 and Figure 16, respectively. The $n$-bit Fredkin gate has $n$–2 input qubits (controls) and two out qubits (switchable targets). Figure 17 illustrates the q-sphere for the composed 3-bit Fredkin gate using the GALA-3 AND operator, when the single input qubit ($in_0$) is set to the states of $|0\rangle$ and $|1\rangle$ for all permutations of the switchable out qubits ($out_0$ and $out_1$). Figure 18(a) and Figure 18(b) demonstrate the q-spheres for the composed 3-bit Miller gate using the GALA-3 AND operator, when the single out qubit is set to the states of $|0\rangle$ and $|1\rangle$ for all permutations of the input qubits ($in_0$ and $in_1$), respectively.

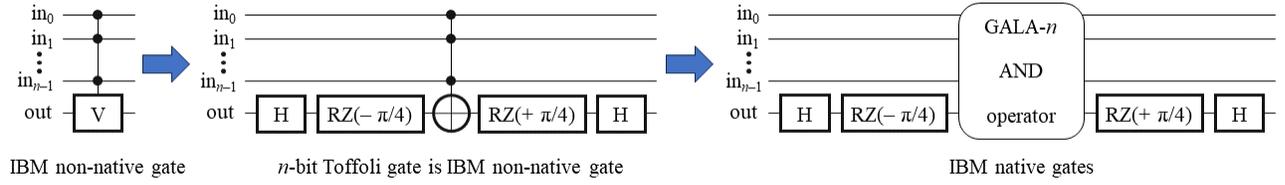

(a)

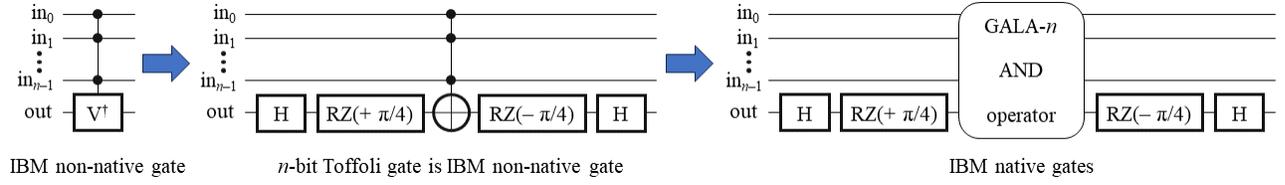

(b)

Figure 14. Schematic of proposed cost-effective quantum gates decomposition from IBM non-native gates to IBM native gates using GALA-$n$ AND operators: (a) the decomposition of $n$-bit controlled-V gate and (b) the decomposition of $n$-bit controlled-$V^\dagger$ gate. The H gates are implicitly decomposed to the sequence {RZ(+ $\pi$/2) $\sqrt{X}$ RZ(+ $\pi$/2)} of IBM native gates.

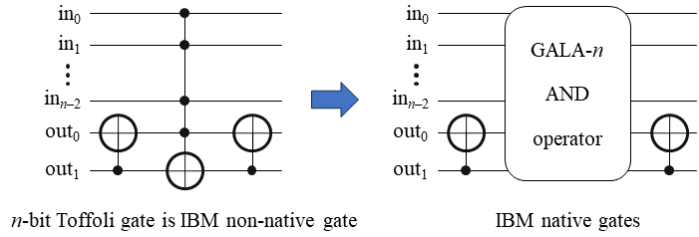

Figure 15. Schematic of the composed $n$-bit Fredkin (multiple-controlled-SWAP) gate using GALA-$n$ AND operator. This composed $n$-bit Fredkin gate has $n$–2 input qubits and two out qubits (out$_0$ and out$_1$).

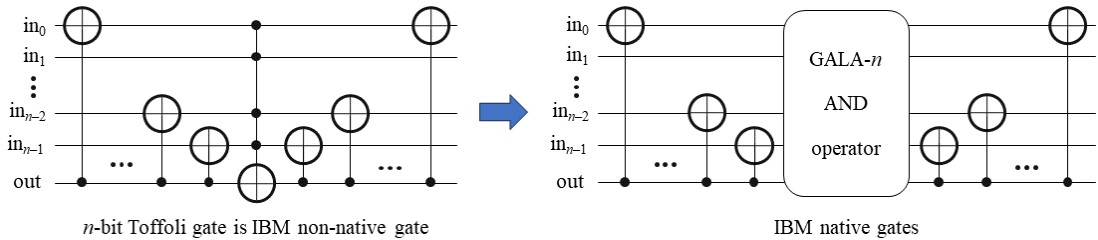

Figure 16. Schematic of the composed $n$-bit Miller gate using GALA-$n$ AND operator. This composed $n$-bit Miller's gate has $n$–1 input qubits and one out qubit.

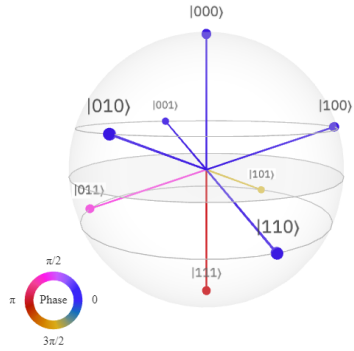

Figure 17. Q-sphere for the outcomes of the composed 3-bit Fredkin gate using the GALA-3 AND operator, when the input qubit ($in_0$) is set to the states of $|0\rangle$ and $|1\rangle$ for all combinations of the out qubits ($out_0$ and $out_1$). For all qubits in this q-sphere, their Dirac notations are in the form of $|out_1\ out_0\ in_0\rangle$.

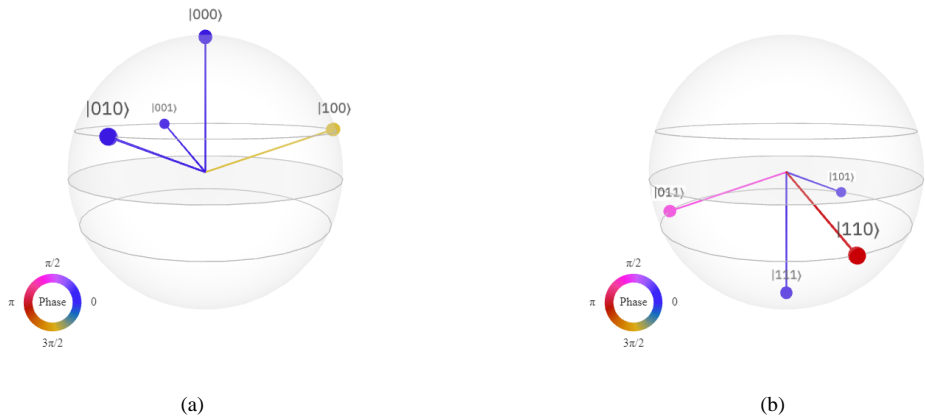

Figure 18. Q-spheres for the outcomes of the composed 3-bit Miller gate using GALA-3 AND operator: (a) the out qubit initially set to the state of $|0\rangle$ for all combinations of input qubits ($in_0$ and $in_1$), and (b) the out qubit initially set to the state of $|1\rangle$ for all combinations of input qubits ($in_0$ and $in_1$). For all qubits in these q-spheres, their Dirac notations are in the form of $|out\ in_1\ in_0\rangle$. Note that when the out qubit is $|0\rangle$ and the input qubits are $|11\rangle$, then the outcome becomes $|100\rangle$ as in (a); and when the out qubit is $|1\rangle$ and the input qubits are $|00\rangle$, then the outcome becomes $|011\rangle$ as in (b).

## 2.4 Transpilation Quantum Cost

From Definition 3 in our work [2], we introduced a new concept of calculating the final quantum cost for a transpiled quantum circuit into the layout of an IBM QPU, which is termed the "transpilation quantum cost (TQC)". The TQC is the sum of the total numbers of (i) IBM native single-qubit gates (I, X, $\sqrt{X}$, and RZ), (ii) IBM native double-qubit gates (CNOT), (iii) the decomposed SWAP gates among the physical qubits of an IBM QPU as shown in Figure 19, and (iv) the depth for the final transpiled quantum circuit.

***Definition 3.*** *The transpilation quantum cost (TQC) of a transpiled quantum circuit into an IBM QPU is formulated as*

$$TQC = N_1 + N_2 + XC + D. \qquad (12)$$

*Where,*
*$N_1$ is the total number of IBM native single-qubit gates (I, X, $\sqrt{X}$, and RZ),*
*$N_2$ is the total number of IBM native double-qubit gates (CNOT),*
*XC is the total number of decomposed SWAP gates among the physical qubits of an IBM QPU, and*
*D is the depth, as the critical longest path of $N_1$, $N_2$, and XC, for the final transpiled quantum circuit* □

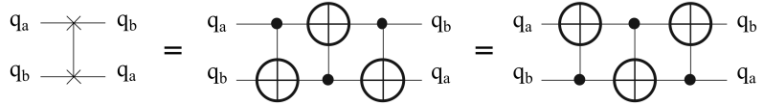

Figure 19. Schematics of IBM non-native SWAP gate (left-side) and its equivalent decomposition of three IBM native CNOT (Feynman) gates (middle and right-side), for swapping arbitrary indexed physical qubits ($q_a$ and $q_b$) of an IBM QPU.

Due to our proposed straight mirrored hierarchical structure of quantum circuits, the TQC of GALA-*n* quantum operators can be predicted from that of GALA-(*n* – 1) quantum operators ahead of the transpilation time, as it will be discussed in Section 4.

IBM quantum system replaced its native CNOT gate with the new native echoed cross-resonance (ECR) gate, to minimize drift in error rates of recalibrations and coherent errors, and to maximize the stability and extend the measuring lengths [10, 11, 41]. For this reason, the $N_2$ and XC parameters, as stated in (12) above, are now counting the total number of native ECR gates among the physical qubits of an IBM QPU.

In this paper, by utilizing the `ibm_brisbane` QPU for $3 \le n \le 4$ qubits, GALA-*n* quantum (Boolean and Phase) operators always have XC = 0, due to the manufactured layout of neighbor physical qubits of this IBM QPU, i.e., the TQC is always lower than that of other conventional *n*-bit quantum (Boolean and Phase) operators. Neighbor physical qubits mean that they are directly connected through channels (buses), and Figure 20 depicts the layout of `ibm_brisbane` QPU of 127 qubits. For the XC and D parameters in (12) above, the TQC of our proposed generic architecture approach mainly depends on the number of neighbor physical qubits of an IBM QPU. Such that, when the number of neighbor physical qubits increases, the TQC of a GALA-*n* quantum operator is dramatically reduced.

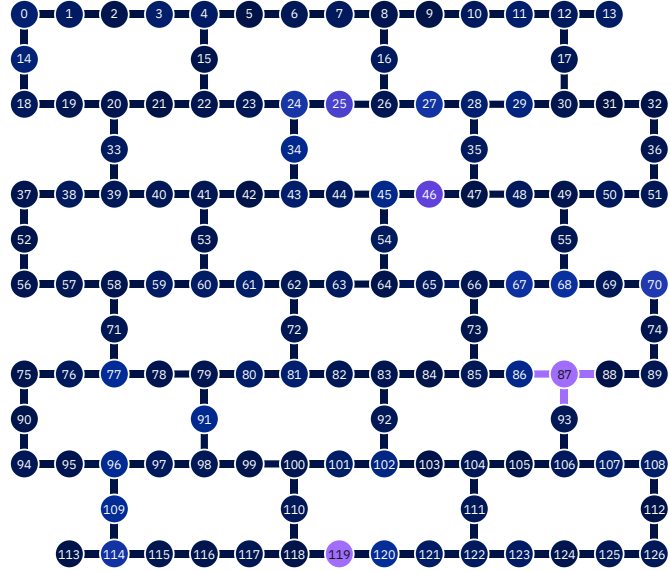

Figure 20. The layout of `ibm_brisbane` QPU of 127 qubits: (i) the colors of physical qubits (circles) indicate the T1 (relaxation time in μsec), T2 (decoherence time in μsec), frequency (in GHz), or readout errors, (ii) the colors of channels (lines) indicate the gates time (in nsec) or ECR errors, and (iii) the circles and lines are scaled from lower (darker color) to higher (lighter color) values in (i) and (ii) [9].

As shown in Figure 20, the physical qubit indexed '0' has two neighbor physical qubits '1' and '14'; such that, the TQC of generic architecture is reduced when $n = 3$ qubits for XC = 0 and fewer D. However, the TQC is increased when $n > 3$ qubits, since more decomposed SWAP gates need to be inserted among the non-neighbor physical qubits, due to the increased values of XC and D parameters. The physical qubit indexed '4' has three neighbor physical qubits '3', '5' and '15'; such that, the TQC of generic architecture is reduced when $3 \leq n \leq 4$ qubits for XC = 0 and fewer D. However, the TQC is increased when $n > 4$ qubits, since more decomposed SWAP gates are inserted for non-neighbor physical qubits.

Hence, due to the recent layouts of IBM QPUs, the minimized (or the optimal) TQC values can be found when $3 \leq n \leq 4$ qubits, and the maximized (or the critical) values of TQC can be found when $n > 4$ qubits. We investigated the optimal and the critical values of TQC as proportional relationships with $n$ neighbor physical qubits based on specific layouts of an IBM QPU, and these values of TQC were termed "optimal-$n$" and "critical-$n$" factors as stated in Definition 4 and Definition 5 in our work [2], respectively.

***Definition 4.*** *The optimal-n parameter defines the effective number of n neighbor physical qubits, to construct the GALA-n quantum operator for an IBM QPU of the layouts (linear, T-like, and I-like), where $3 \leq n \leq 4$ qubits. The optimal-n parameter ensures to (i) have any three (or four) adjacent physical qubits being directly connected through two (or three) channels, (ii) minimize the transpilation quantum cost (TQC), and (iii) maintain the straight mirrored hierarchical structure for the final transpiled quantum circuit □*

***Definition 5.*** *The critical-n parameter defines an arbitrary number of n physical qubits, to construct the GALA-n quantum operator for an IBM QPU of the layouts (linear, T-like, and I-like), where $3 \leq n \leq$ total number of physical qubits. In the critical-n parameter, (i) arbitrary n physical qubits are not (or mostly not) directly connected through channels, (ii) a GALA-n quantum operator may lose its straight mirrored hierarchical structure, (iii) SWAP gates may be added among the physical qubits, and (iv) the transpilation quantum cost (TQC) is dramatically increased as n increases □*

For instance, for a lower TQC of the final transpiled quantum circuit using `ibm_brisbane` QPU, Figure 21 depicts the construction of different GALA-4 quantum operators and composed gates, where all their input qubits are mapped into the physical qubits of indices '3', '5', and '15', and their out qubits are mapped into the physical qubit of index '4'. However, for the composed 4-bit Fredkin gate, its two input qubits are mapped to the physical qubits of indices '5' and '15', and its two out qubits are mapped into the physical qubits of indices '3' and '4'.

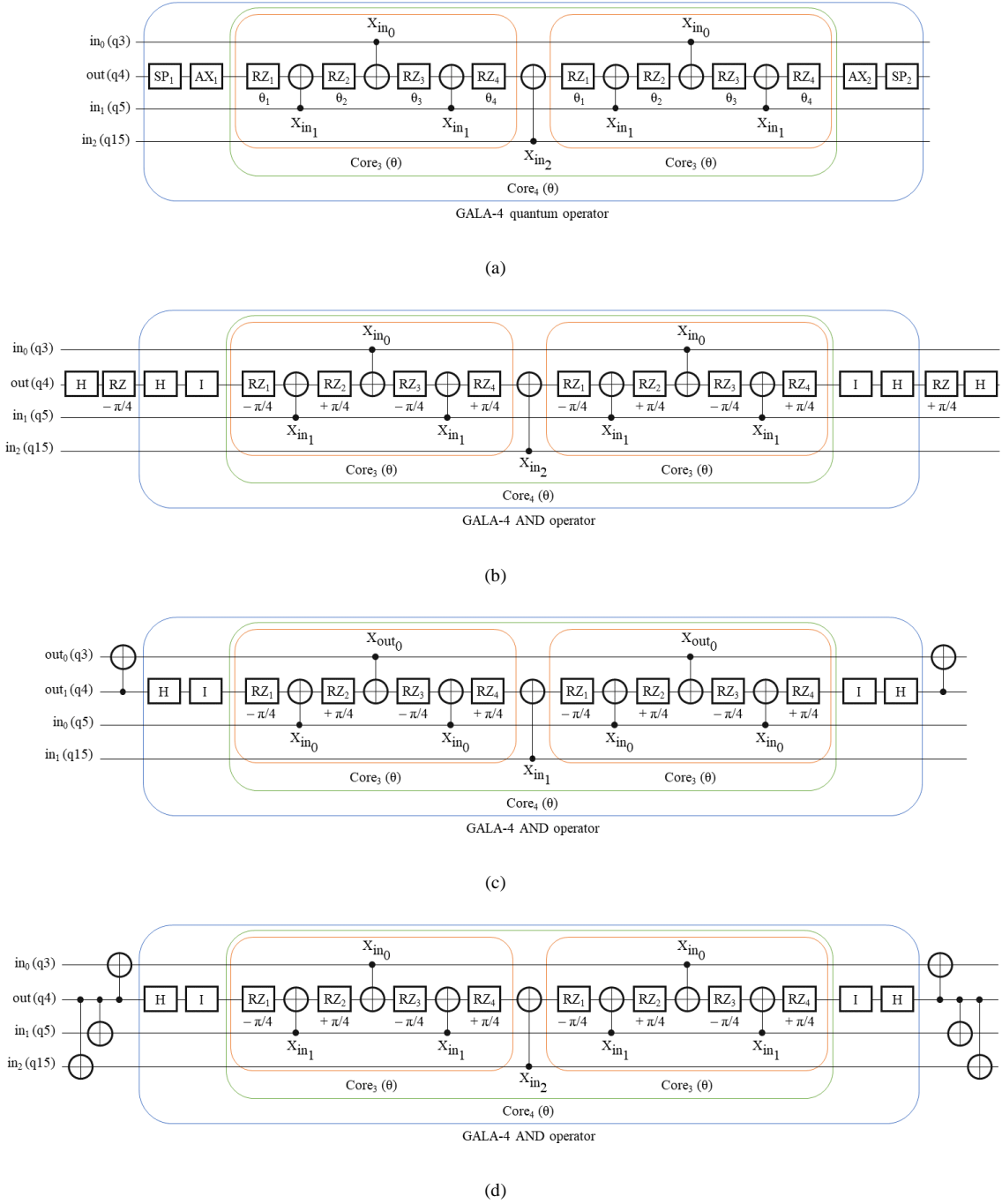

Figure 21. Construction of different GALA-4 quantum operators and composed gates based on the layout of `ibm_brisbane` QPU: (a) GALA-4 quantum (Boolean or Phase) operator based on Table 1, (b) the composed 4-bit controlled-V gate, (c) the composed 4-bit Fredkin gate, and (d) the composed 4-bit Miller gate. All the composed gates use GALA-4 AND operators. For a lower TQC of the final transpiled quantum circuits, the input and out qubits are mapped to q3, q4, q5, and q15 properly. The q3, q4, q5, and q15 are the physical qubits of `ibm_brisbane` QPU for the indices 3, 4, 5, and 15, respectively.

## 3 CASE STUDIES AND RESULTS

Different experiments of the conventional *n*-bit quantum operators and GALA-*n* quantum operators are built, transpiled using `ibm_brisbane` QPU of 127 qubits, and then their cost-effectiveness is evaluated based on our proposed transpilation quantum cost (TQC), as stated in (12) above. In this paper, the experiments are exercised and evaluated for $3 \leq n \leq 4$ qubits, since our basic IBM Quantum account (Open Plan) has limited execution jobs per month [42].

The conventional *n*-bit quantum operators are built using conventional non-native quantum gates, such as *n*-bit Toffoli gate, GALA-*n* quantum operators are generated using ALGORITHM 1, and GALA-based composed gates are constructed as earlier discussed in Section 2.3 and Section 2.4. The inputs (controls) and outputs (targets) of conventional *n*-bit quantum operators are randomly assigned to the physical qubits of `ibm_brisbane` QPU, based on the IBM transpilation software (Transpiler). However, the inputs and outputs of GALA-*n* quantum operators are specifically assigned to their equivalent physical qubits of `ibm_brisbane` QPU, based on ALGORITHM 1 that strictly depends on the defined indices of the physical qubits and the geometrical layout of this IBM QPU. Due to these physical qubits' assignments, the Transpiler always inserts more SWAP (as three ECR) gates among the physical qubits for the conventional *n*-bit quantum operators than that of GALA-*n* quantum operators. For this reason, the transpiled GALA-*n* quantum operators always have lower TQC than those of the transpiled conventional *n*-bit quantum operators.

The IBM Quantum Platform was utilized to design and examine these experiments for 1024 shots [43]. However, the technical specifications of T1, T2, frequency, anharmonicity, median errors of the native gate, speed of transpilation, number of required pulses, speed of Internet connection, usage load of `ibm_brisbane` QPU are not considered for the purpose of our research and experiments.

To save space, the final transpiled quantum circuits of the conventional *n*-bit quantum operators and GALA-*n* quantum operators are not demonstrated in this section. However, all constructed and transpiled quantum circuits with their relevant data are available upon request from the authors. To illustrate the purpose of this research, Figure 22 depicts the constructed quantum circuits of conventional 3-bit and 4-bit quantum operators, and their equivalent cost-effective GALA-3 and GALA-4 quantum operators are constructed as previously discussed in Section 2.2, Section 2.3, and Section 2.4, with their inputs and out qubits are specifically mapped to the physical qubit of indices '3', '4', '5', and '15', as previously shown in Figure 20.

Table 3 summarizes the calculations of our proposed transpilation quantum cost (TQC), in (12) above, for the final transpiled quantum circuits of different experiments (as conventional *n*-bit quantum operators compared with GALA-*n* quantum operators), using `ibm_brisbane` QPU, where $3 \leq n \leq 4$ qubits.

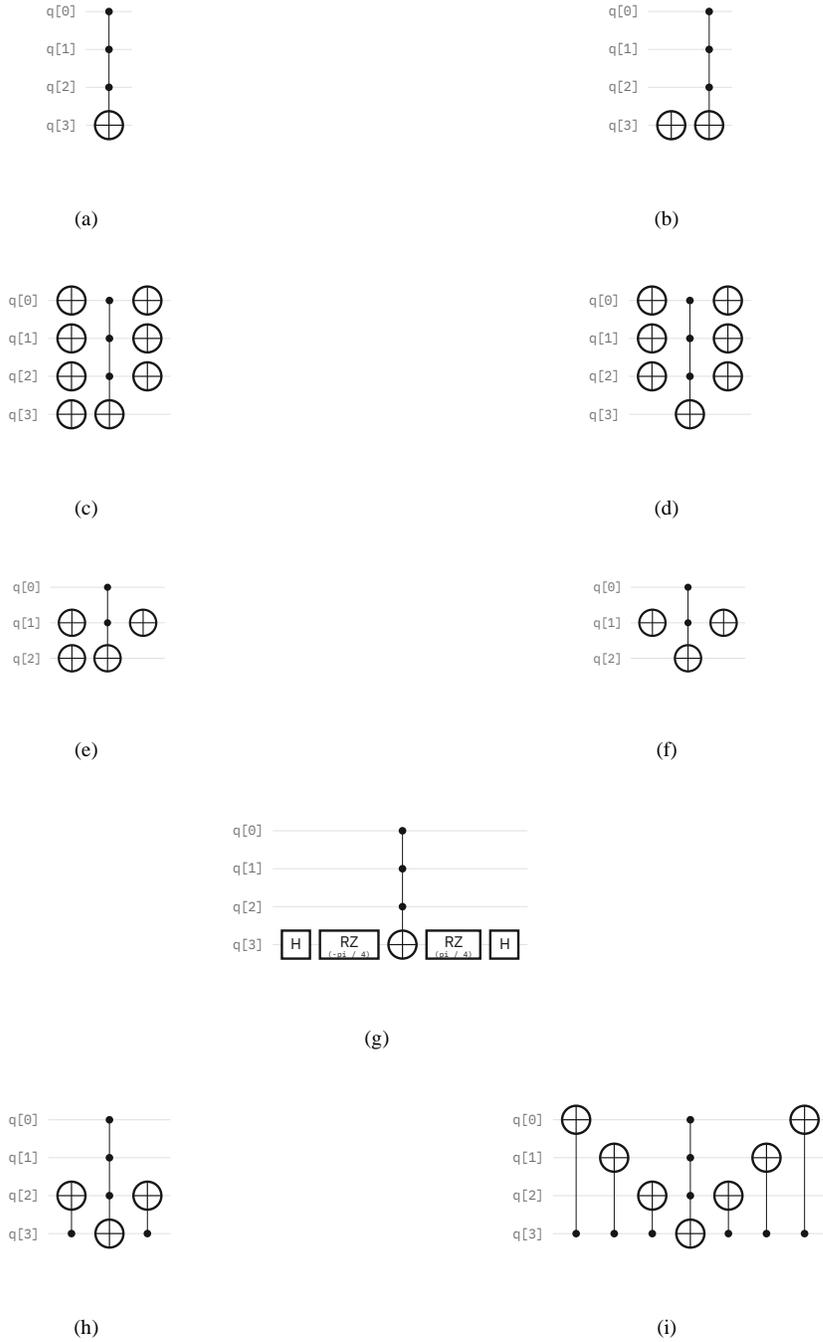

Figure 22. Schematic of the constructed quantum circuits for the conventional quantum operators: (a) 4-bit AND, (b) 4-bit NAND, (c) 4-bit OR, (d) 4-bit NOR, (e) 3-bit Implication, (f) 3-bit Inhibition, (g) 4-bit Controlled-V, (h) 4-bit Fredkin, and (i) 4-bit Miller. The q (inputs and outputs) are arbitrary qubits and not the physical qubits of `ibm_brisbane` QPU. All these quantum circuits are not transpiled yet.

Table 3. Summary of TQC for the transpiled quantum circuits for the conventional *n*-bit quantum operators and GALA-*n* quantum operators using `ibm_brisbane` QPU, where $3 \leq n \leq 4$ qubits.

| *n*-bit quantum operator | *n* qubits (inputs + out) | Transpiled quantum circuit for conventional *n*-bit approach | | | | | Transpiled quantum circuit for GALA-*n* approach | | | | |
|---|---|---|---|---|---|---|---|---|---|---|---|
| | | $N_1$ | $N_2$ | XC | D | TQC | $N_1$ | $N_2$ | XC | D | TQC |
| AND | 4 | 146 | 29 | 5 | 108 | **288** | 52 | 7 | 0 | 41 | **100** |
| NAND | 4 | 105 | 20 | 2 | 83 | **210** | 52 | 7 | 0 | 41 | **100** |
| OR | 4 | 102 | 20 | 2 | 80 | **204** | 52 | 7 | 0 | 41 | **100** |
| NOR | 4 | 103 | 20 | 2 | 83 | **208** | 52 | 7 | 0 | 41 | **100** |
| Implication | 3 | 49 | 9 | 1 | 39 | **98** | 28 | 3 | 0 | 21 | **52** |
| Inhibition | 3 | 50 | 9 | 1 | 39 | **99** | 28 | 3 | 0 | 21 | **52** |
| Controlled-V | 4 | 147 | 29 | 5 | 110 | **291** | 53 | 7 | 0 | 42 | **102** |
| Fredkin | 4 | 189 | 40 | 8 | 115 | **352** | 56 | 9 | 0 | 46 | **111** |
| Miller | 4 | 151 | 29 | 3 | 113 | **296** | 70 | 13 | 0 | 58 | **141** |

From Table 3, the following key points were observed when using `ibm_brisbane` QPU for $3 \leq n \leq 4$ qubits:

1. The TQC for all conventional *n*-bit quantum operators has unpredictable values, since the inputs and out qubits of these operators are randomly assigned (by Transpiler) to the physical qubits of this QPU. Such that, the TQC may significantly increase or decrease for different transpilation attempts!
2. The TQC for all GALA-*n* quantum operators has predictable values, since the inputs and out qubits of these operators are specifically assigned (by ALGORITHM 1) to the physical qubits of this QPU, before Transpiler. Such that, the TQC is always fixed for different transpilation attempts.
3. The TQC for all conventional *n*-bit quantum operators is at least twice as high as that of GALA-*n* quantum operators, in which GALA-*n* quantum operators become more generic cost-effective realized quantum circuits for IBM QPUs.
4. For all GALA-*n* quantum operators, the XC parameter always equals 0, due to the straight structure of their constructed quantum circuits.
5. For all GALA-4 quantum operators, the $N_1$, $N_2$, and D parameters are approximately doubled from those of GALA-3 quantum operators, due to the straight mirrored structure of their constructed quantum circuits.
6. For all composed GALA-4 quantum operators, the TQC for controlled-V, Fredkin, and Miller gates is slightly higher than that of GALA-4 AND operator, due to the straight mirrored hierarchical structure of their constructed quantum circuits with the added functional (extra) quantum gates, e.g., the CNOT gates.

When $n > 4$ qubits, from the aforementioned key points, the TQC of GALA-*n* quantum operators can be predicted as a slightly doubled value from that of GALA-($n - 1$) quantum operators, due to the straight mirrored hierarchical structure of their quantum circuits as well as due to the increased value of XC parameter. Hence, the prediction of TQC is another interesting feature of GALA-*n* quantum operators. This interesting feature is mainly useful for cost-effective quantum circuits realization into IBM QPUs ahead of the transpilation time.

It is concluded that GALA-*n* quantum operators (as ALGORITHM 1) and the composed GALA-*n* quantum operators (as an expansion to ALGORITHM 1) can be both employed as transpilation libraries for the IBM quantum system, since their transpiled quantum circuits always have much lower TQC than that of the transpiled conventional *n*-bit quantum operators.

## 4 CONCLUSION

The conventional *n*-bit Toffoli gate is mainly used to construct different conventional *n*-bit quantum operators, for both Boolean and Phase operations, where $n \geq 3$ qubits. A conventional *n*-bit quantum operator is a set of non-native gates for IBM quantum processing units (QPUs). For practical quantum applications, this set is decomposed into another set of IBM native gates to be mapped properly into the layout (or topology) of an IBM QPU. In IBM terminologies, the decomposition and mapping mechanisms belong to the transpilation process. In this paper, our research mainly focuses on the cost-effective transpilation for different conventional *n*-bit quantum operators using `ibm_brisbane` QPU, where $3 \leq n \leq 4$ qubits. After transpilation, we observed the following disadvantageous outcomes for the final transpiled quantum circuits: (i) a large decomposition set of IBM native gates is generated, (ii) too many decomposed SWAP gates are added among the physical qubits of this QPU, and (iii) the depth (as the critical longest path along the native gates) is highly increased. Hence, the final quantum cost of these transpiled quantum circuits is dramatically increased. For that, we designed a cost-effective generic architecture of layout-aware *n*-bit quantum operators, which are termed "GALA-*n* quantum operators", for both Boolean and Phase operations and $n \geq 3$ qubits. The quantum circuit of the GALA-*n* quantum operator is designed as a straight mirrored hierarchical structure of: (i) a small set of IBM native single-qubit gates (X, $\sqrt{X}$, and RZ) and IBM native double-qubit gates (CNOT or ECR), (ii) fewer decomposed SWAP gates among the physical qubits, and (iii) a lower depth.

The construction of GALA-*n* quantum operators mainly depends on the layout and the number of *n* neighbor physical qubits of an IBM QPU. GALA-*n* quantum operators are entirely designed using the visual approach of the Bloch sphere, from the visual representations of the rotational quantum operations for IBM native (X, $\sqrt{X}$, RZ, and CNOT) gates. Moreover, we proposed a new formula for calculating the quantum cost for the final transpiled quantum circuits, and this formula is termed the "transpilation quantum cost (TQC)". The TQC calculates the total numbers of (i) IBM native single-qubit gates, (ii) IBM native double-qubit gates, (iii) the decomposed SWAP gates, and (iv) the depth. After transpilation using `ibm_brisbane` QPU for $3 \leq n \leq 4$ qubits, our proposed GALA-*n* quantum operators always have lower TQC than that of the conventional *n*-bit quantum operators. Different experiments were examined and evaluated, including GALA-*n* AND, NAND, OR, NOR, Implication, and Inhibition operators and the composed GALA-*n* controlled-V, controlled-V†, Fredkin, and Miller gates, and compared with the conventional *n*-bit AND, NAND, OR, NOR, implication, inhibition, controlled-V, controlled-V†, Fredkin, and Miller gates. It is concluded that: (i) the TQC for all conventional *n*-bit quantum operators is at least twice as high than that of GALA-*n* quantum operators, (ii) the TQC for all conventional *n*-bit quantum operators has unpredictable values, (iii) the TQC for all GALA-*n* quantum operators has predictable values, and (iv) GALA-*n* quantum operators are more generic cost-effective realized quantum circuits for IBM QPUs. The TQC of GALA-*n* quantum operators can be easily predicted from that of GALA-($n$ – 1) quantum operators ahead of the transpilation time, and such a prediction of TQC is an interesting feature for cost-effective quantum circuits realization into IBM QPUs in advance. In this paper, we introduce GALA-*n* quantum operators as introductory research for building cost-effective Boolean and Phase operators for quantum oracles, quantum search algorithms, and future quantum computing enhancements for efficient mapping to different layouts of IBM QPUs. Accordingly, GALA-*n* quantum operators and the composed GALA-*n* quantum operators are proposed as transpilation libraries for the IBM quantum system, due to the significantly lower TQC of their final transpiled quantum circuits.